\documentstyle[12pt,epsfig]{article}
\def\ifig#1#2#3#4{\begin{figure}[hbtp]
 \begin{center}\leavevmode\epsfig{file=#1,height=#2}
 \caption[#3]{#4}\end{center}\end{figure}}

\title{          On the Shell Structure of Nuclear Bubbles
    \footnote{This work is partly  supported by the Polish Committee of
          Scientific Research under contract No.  2P03B01112}}

\author{     Klaus Dietrich and Krzysztof Pomorski
    \thanks{On leave on absence from University M.C.S. in Lublin}\\
      Technische Universit\"at M\"unchen, Garching, Germany}

\begin{document}

\maketitle
          
\begin{abstract}
We investigate the shell structure of spherical nuclear bubbles
in simple phenomenological shell model potentials. The bunching
is produced by the energy gaps between single-particle states
with a~different number of radial nodes.  The shell correction
energies for doubly magic bubbles may be as large as --40~MeV
and probably imply a~very long lifetime against spontaneous fission.
$\beta$-stability occurs for ratios of the neutron number
$N$ to the proton number $Z$ which differ markable from the
$\beta$-stability valley of ordinary compact nuclei. 
The $\alpha$--decay probability is shown to be very small for 
proton rich bubbles with a moderately large outer radius.
Metastable islands of nuclear bubbles are shown to exist for 
nucleon in the range $A$=450--3000.
\end{abstract}

\section{Introduction}

The idea that stable or metastable nuclear bubbles might exist
is rather old. Already in 1946 Wilson \cite{Wi46} has written
about spherical bubble nuclei.
J.A. Wheeler \cite{Whee}, in his famous notebook, mention the
possibility that nuclei with different topology might exist.
In 1967, P.J. Siemens and H.A. Bethe \cite{Si67} studied the
energy of strongly deformed spheroidal nuclei and of spherical
bubble nuclei using a liquid drop model (LDM). They looked at
nuclei in the region of the then expected superheavy nuclei but
did not investigate the stability of the spherical bubbles 
with respect to deformations.
F. Wong \cite{Wo73} was the first to perform detailed studies:
Within the pure LDM he showed that for a fissibility parameter $X_0 > 2$
spherical nuclear bubbles have a lower energy than compact spherical
nuclei but are not stable against fission. He then investigated the shell 
correction energy for toroidal and spherical bubble nuclei
on the basis of a harmonic oscillator potential
using the Strutinsky's method. In the study of the shell correction energy,
he restricted his attention to known $\beta$--stable nuclei and 
found spherical bubbles with a very small inner radius as lowest
energy configurations for certain doubly-- magic known nuclei ($^{138}Ce$,
$^{200}Hg$).

More recently, L. Moretto et al. \cite{Mo97} investigated the effect
of an ideal dilute classical gas of nucleons at finite temperature 
inside the bubble and claimed that the pressure exerted by this gas
on the layer of nuclear matter could stabilize
spherical bubbles against deformation.  As the authors pointed out,
this mechanism may also be of relevance for the existence 
of ''mesoscopic bubbles''.

In what follows we investigate the effects of the shell
structure for spherical bubbles in a large range of neutron ($N$)
and proton ($Z$) numbers considerably extending them  beyond the known
nuclear species. The present paper is an extension of our study presented
in Ref. \cite{Di97}.  We use phenomenological shell
model potentials and apply the Strutinsky method to 
evaluate the shell correction energy $\delta E_{shell}$. The
motivation of this work is to find the magic neutron 
and proton numbers and to obtain an estimate of 
the magnitude of the shell energy. This simple 
approach is expected to be sufficiently
realistic to establish the importance
of shell effects in nuclear bubbles and to lead to 
a~first assessment of the stability of such systems.

Surely, a~reliable investigation of the existence of
nuclear bubbles with an appreciable life-time
versus fission and other decay channels requires
careful and technically difficult Hartree-Fock (HF)
or Hartree-Bogoliubov (HB) calculations.  Indications
that bubble solutions may exist have indeed
already been found recently in HB-calculations
with Gogny interactions \cite{De97} and some time ago \cite{Bo97}
in HF-calculations based on the Skyrme interaction.

\section{Theory}

As in the original shell model of M. G\"oppert-Mayer
and J. H. D. Jensen \cite{Gop}, we investigate the
shell structure in the two limiting cases of a~square
well with infinite walls, on the one hand,
and an oscillator potential, on the other.

In order to incorporate the case that the interior of
the bubble is not entirely empty but filled
by an internal halo of nucleons, we also
investigate the effect of replacing the above
mentioned single particle potentials by a~constant
potential of (small) finite depth $V_2$
for radial coordinates smaller than the inner radius ($R_2$)
of the bubble.

The spin-orbit potential
for a bubble nucleus is smaller than for
an ordinary compact nucleus because the contributions
arising from the inner and outer surface
have a~different sign. Therefore, we treated
the spin-orbit potential in perturbation theory.

\subsection{Eigenfunctions and eigenenergies of the spin-independent 
 potentials}

In the coupled representation which will turn out to be useful for the
perturbative treatment of the spin--orbit potential, the eigenstates
of the unperturbed Hamiltonian
$$
 \widehat H_0 = - {\hbar^2\over 2M} \Delta + V(r) 
\eqno(2.1)
$$
have the form
$$
 \psi_{nljm} (r,\theta,\phi) = \varphi_{nl}(r) {\cal Y}_{ljm} 
 (\vartheta,\varphi)
\eqno(2.2)
$$
$$
 {\cal Y}_{ljm} : = \sum_{m_l,m_s} \langle lm_l{1\over 2}m_s |
  l{1\over 2}jm\rangle \, Y_{lm_l} (\theta,\phi) \chi_{m_s}
\eqno(2.3)
$$
Here, $V(r)$ is a~phenomenological spherically symmetric shell
model potential. The quantities appearing in (2.3) have an
obvious meaning. The radial wave functions $\varphi_{nl}(r)$ are
normalized eigenfunctions of the radial Schr\"odinger equation
$$
 -{\hbar^2\over 2M} \left({\partial^2\over \partial r^2} + 
  {2\over r} {\partial\over \partial r} - {l(l+1) \over
  r^2}\right) \varphi_{nl} + \left(V(r) - \buildrel\over\varepsilon_{nl}
  \right) \varphi_{nl}(r) = 0\,,
\eqno(2.4)
$$
which have to satisfy the boundary conditions 
$$
 \varphi_{nl}(0) = 0 ~~~~~~~\mbox{for all}~~~l\neq 0;
   ~~~\lim_{r\rightarrow 0} (r\varphi_{n0}(r))=0 \;\;,\\
\eqno(2.5.1)
$$
$$
 \varphi_{nl}(\infty) = 0 ~~~~~~\mbox{for all}~~~l~~~~~~~~~~~ \;\;.
\eqno(2.5.2)
$$
The eigenvalues $\buildrel\over\varepsilon_{nl}$ and
eigenfunctions $\varphi_{nl}(r)$ depend on the orbital angular
momentum $l$ and on the number of radial nodes. Not counting the
zeros at $r=0,\infty$, the number of radial nodes is given by ($n-1$)
where $n=1,2,...$ .

\noindent
{\bf i)} In the case of the infinite square-well 
$$
 V(r) = 
 \left\{
 \begin{array}{ll}
 -V_0 & \mbox{for}~~~R_2 < r < R_1 \\
 +\infty & \mbox{otherwise}
\end{array}\right.
\eqno(2.6)
$$
the boundary conditions (2.5) are replaced by
$$
 \varphi_{nl}(R_2) = \varphi_{nl}(R_1) = 0\,.
\eqno(2.7)
$$
A finite well depth $V_0$ (= 50 MeV) is introduced for later
convenience (see Eq. (2.11)). It has no influence on the shell 
correction energy.
The eigenstates $\varphi_{nl}(r)$ are linear combinations of
a~spherical Bessel function $j_l(\alpha r)$ and a~spherical
Neumann function $y_l(\alpha r)$ (for ''spherical Bessel function 
of the first and second kind'', see Ref. \cite{HMF})
$$
 \varphi_{nl}(r) = N_{nl} [j_l(\alpha r) + b_{nl} \, y_l(\alpha r)]\,.
\eqno(2.8)
$$
The parameter $\alpha$ is defined by
$$
 \alpha = \sqrt{{2M(\varepsilon_{nl} + V_0)\over\hbar^2}}
\eqno(2.9)
$$
and the normalization constants $N_{nl}$ are to be determined by
the condition
$$
 \int dr \, r^2 |\varphi_{nl}(r)|^2 = 1\,.
\eqno(2.10)
$$
The two parameters $\varepsilon_{nl}$ and $b_{nl}$ are obtained
by satisfying numerically the boundary conditions (2.7).

\medskip
\noindent
{\bf ii)} In order to describe a finite density in the interior ($r <
R_2$) of the bubble, we consider the slightly more complicated
potential 
$$
 V(r) =
 \left\{\begin{array}{lll}
 -V_2 & {\rm for} & 0 \leq r < R_2 \\
 -V_0 & {\rm for} & R_2 < r < R_1 \\
 +\infty & {\rm for} & r > R_1 
 \end{array}\right.\;\;,
\eqno(2.11)
$$
where the depth parameter $V_2$ ($>0$) in the interior region
is substantially smaller than $V_0$ ($>0$). As a~rough estimate one may
choose the ratio ${V_2/V_0}$ to be equal to the ratio
$\rho^{(2)}_{n(p)}/\buildrel\circ\over\rho_{n(p)}$ of the average 
density $\rho^{(2)}_{n(p)}$  of neutrons (protons) in the region
$r < R_2$ to the uniform LD density $\buildrel\circ\over\rho_{n(p)}$
of neutrons (protons) in the region $R_2 < r < R_1$. The radial
functions $\varphi_{nl}(r)$ and the eigenenergies 
$\buildrel\circ\over\varepsilon_{nl}$ which correspond to the 
infinite square well with internal step (2.11) are given in
Appendix A1.

\medskip
\noindent
{\bf iii)} As an opposite extreme to the one of the infinite square
well we consider a~harmonic oscillator with a~finite internal step:
$$
 V(r) = 
 \left\{
 \begin{array}{ll}
 -V_2 & {\rm for}~~0 \leq r < R_2~~{\rm (region~~II)} \\
-V_0 + {M\omega^2\over 2}(r - \bar R)^2 & {\rm for}~~r > R_2~~~~~
 {\rm (region~~I)}
\end{array}\right.\,,
\eqno(2.12)
$$
where
$$
\bar{R} := \frac{1}{2} (R_1 + R_2) \,\,.
\eqno(2.13)
$$
In Appendix A2 we exhibit the mathematical form of the radial 
eigenfunctions $\varphi_{nl}(r)$ for the potential (2.12) and
discuss the numerical methods which may serve to obtain them.

The WKB method is known to yield the eigenvalues of simple nuclear 
potential models rather well we have therefore applied it in our case.

The upper (b) and lower (a) turning points for a~particle
of energy $\varepsilon$ and orbital angular momentum $l$ in the 
potential (2.12) are given as the real solutions of the equation 
$$
 \varepsilon = {\hbar^2 l(l+1)\over 2Ma^2} + V(a) = {\hbar^2l(l+1)\over
     2Mb^2} + V(b)\,.
\eqno(2.14)
$$
As the analytic roots of this quartic equations are complicated,
it is more convenient to solve (2.14) numerically. The
WKB-eigenvalues $\varepsilon_{nl}$ are defined as 
the solutions of the equation \cite{Mes}
$$
 \sqrt{2M} \int^b_a dr \sqrt{\varepsilon - V(r) - 
 {\hbar^2 l(l+1)\over 2Mr^2}} = \left(n + {1\over 2}\right) \hbar\pi\,.
\eqno(2.15)
$$
Here again an analytical integration is possible but
impracticable whereas the numerical integration can be easily
performed.

So far, we neglected the average Coulomb potential $V_{\rm
Cb}(r)$ acting on the protons
$$
V_{\rm Cb} (r) = e_0 \int d^3r' {\rho^{(p)}(\vec {r'})\over 
                 |\vec r - \vec {r'}|}\,.
\eqno(2.16)
$$
If one replaces the density distribution $\rho^{(p)}(\vec r)$
of the protons by a uniform distribution in the layer between
the inner and outer LD-radius
$$
\rho^{(p)}(\vec r) = \Theta_0(r-R_2) \Theta_0(R_1-r)
                      {\buildrel\circ\over\rho}_p  \,\,,
\eqno(2.17)
$$
$$
 \buildrel\circ\over\rho_p = {3 Z e_0 \over 4\pi (R^3_1 - R^3_2)}\,\,,
\eqno(2.18)
$$
the Coulomb potential $V_{\rm Cb}(r)$ has the form
$$
\begin{array}{ll}
V_{\rm Cb}(r)  &=  {4\pi\over 3} e_0 \buildrel\circ\over\rho_p 
  \left\{\theta_0 (R_2 - r) {3(R^2_1 - R^2_2)\over 2} + \theta_0 
  (r - R_1) \cdot {(R^3_1 - R^3_2)\over r}  \right. \nonumber \\
   &+\, \left. \theta_0 (r - R_2)\theta_0 (R_1 - r) \left({3R^2_1\over 2} -
    {R^3_2 \over r} - {r^2\over 2}\right)\right\} \,\,.
\end{array}
\eqno(2.19)
$$
If one wishes to take this potential into account, the potential
$V(r)$ must be replaced by $[V(r) + V_{\rm Cb}(r)]$ whenever it acts
on protons. 

We draw attention to the well-known result that a proton in the
interior of the bubble ($r < R_2$) feels a constant
Coulomb potential. This means that only the repulsive Coulomb
force between protons in the interior region
disfavours the formation of an internal proton distribution as
compared to the formation of an internal neutron distribution.

At this point, we would like to make the following remark: The
shell correction energy $\delta E_{shell}$ is invariant with respect
to a~constant shift of all the eigenvalues of a~given sort of
particles. The shifts can be different for neutrons and protons.
This is the reason why we may identify the nuclear part $V(r)$
for $n$ and $p$ as long as we neglect the Coulomb potential and
assume that the other parameters
of the potential (radius parameter $R_{1,2}$, frequency $\omega$) 
are the same for $n$ and $p$. This
simple assumption turned out to be empirically very successful
for the Nilsson potential.  Therefore, we adopt it too,
realizing at the same time that it may be insufficiently
accurate for cases of very exotic nuclear compositions.

The effect of a finite potential $V_2 \neq 0$ in the interior region 
$r < R_2$ is qualitatively the same for the two potentials (2.11) and (2.12).
Therefore, we studied the effect of a finite value of $V_2$ only for the 
infinite square well. In all calculations with the harmonic oscillator
potential (2.12) the parameter $V_2$ was chosen to be 0 :
$$
V(r) = -V_0 + {M\omega^2 \over 2} (r - \bar R)^2 {\rm ~~~for~~~} 
  r \geq 0 \,\,.
\eqno(2.20)
$$
Furthermore, we determined the eigenvalues $\varepsilon_{nl}$ 
corresponding to the potential (2.20) from the WKB approximation 
and the radial s.p. densities
$$
\rho_{nl}(r) = \varphi^2_{nl}(r)
$$
from the Thomas-Fermi model. The details of this procedure are
presented in Appendix A3.

We still have to state how the ''shape parameters'' of the shell
model potentials are to be chosen: Since we use the Strutinsky
method representing the total binding energy of the nucleus
as a sum of the liquid drop (LD) energy $E_{\rm LD}$ and the
shell correction energy $\delta E_{\rm shell}$, the shape parameters of
the shell model potential must be chosen in accordance with the
ones of the liquid drop. When calculating the energy as a
function of the nuclear shape, the multipole moments (of low
order) calculated with the shell-model wavefunction must agree
with the ones calculated with the LD-density. For the case of
spherical bubble nuclei, the MS deviation from the central
sphere of radius $\bar R$ must be the same when calculated with
the shell-model wave function and with the LD-density
$$
\langle(r - \bar R)^2\rangle_{\rm SM} = 
  \langle(r - \bar R)^2\rangle_{\rm LD} \,,
\eqno(2.21)
$$
where
$$
\bar R: = {1\over 2}(R_1 + R_2)\,.
\eqno(2.21')
$$
We restrict our consideration to bubble nuclei with a negligible
nuclear density inside of the smaller radius $R_2$ of
the LD-distribution. This means that we only consider the cases
of the infinite square well and of the shifted oscillator without
an internal step (see Eq. (2.20)).

For the shifted oscillator, the condition (2.21) may serve to fix
the frequency $\omega$. For the square well potential it is
physically reasonable to choose the potential walls at a
slightly larger spacing than the difference of the LD-radii
$$
 \tilde R_1 = R_1 + \Delta R \,\,,
\eqno(2.22)
$$
$$
 \tilde R_2 = R_2 + \Delta R \,\,.
\eqno(2.22')
$$
The condition (2.21) may then serve to determine the parameter
$\Delta R$.
One could modify these constraints slightly. So the shift (2.22)
of the inner and outer radius of the potential with respect to
the ones of the liquid drop need not be equal and the reference
radius $\bar R$ (see (2.21')) might be chosen differently. We
believe that the simple choices we made are reasonable
given the fact that the Strutinsky method itself is of
limited accuracy.

The MS-radius calculated within the LDM is given by
$$
\langle(r - \bar R)^2\rangle_{\rm LD} = 
  {{1\over 10}({R_1}^5-{R_2}^5) + {1\over 4}(-{R_1}^4
   R_2 + {R_1}^3 {R_2}^2 - {R_1}^2 {R_2}^3 + R_1{R_2}^4)
  \over ({R_1}^3 - {R_2}^3)}\,.
\eqno(2.23)
$$
The MS-radius $\langle(r - \bar R)^2\rangle_{\rm SM}$ in the
shell model is approximately calculated in the Thomas-Fermi
approximation (see Appendix A3): 
$$
\begin{array}{ll}
\langle (r - \bar R)^2\rangle_{\rm SM} &\approx  {2\over h^3} 
 \int d^3 r \int d^3 p \,(r - \bar R)^2 \left\{\theta_0 \left[\varepsilon^n_F
 - {p2\over 2M} - V(r)\right] + \right. \nonumber \\
 &+ \left. \theta_0 \left[\varepsilon^p_F - {p^2\over 2M} -
V(r) - V_{\rm Cb}(r)\right]\right\}\,.
\end{array}
\eqno(2.24)
$$
The Fermi energies $\varepsilon^n_F$, $\varepsilon^p_F$ for the
neutrons and protons are determined by the number $N$ of
neutrons and $Z$ of protons 
$$
\begin{array}{ll}
N &= {2\over h^3} \int d^3 r\, \int d^3 p \, \theta_0 
  \left[\varepsilon^n_F - {p^2\over 2M} - V(r)\right] \nonumber \\
  &= {32\pi^2(2M)^{3/2}\over h^3} \int^{r^n_1}_{r^n_2} [\varepsilon^n_F
  - V(r)]^{3/2} \,\,,
\end{array}
\eqno(2.25.1)
$$
$$
\begin{array}{ll}
Z &= {2\over h^3} \int d^3 r\, \int d^3 p \, \theta_0 \left[\varepsilon^p_F 
     - {p^2\over 2M} - V(r) - V_{\rm Cb}(r)\right] \nonumber \\
  &= {32\pi^2(2M)^{3/2}\over h^3} \int^{r^p_1}_{r^p_2} [\varepsilon^p_F
     - V(r) - V_{\rm Cb}(r)]^{3/2}\,\,.
\end{array}
\eqno(2.25.2)
$$
All integrations but one can be trivially done and we obtain
$$
\langle(r - R)^2\rangle_{\rm SM} \approx {32\pi^2(2M)^{3\over 2}\over h^3}
(J_n + J_p) \,\,,
\eqno(2.26.1)
$$
$$
J_n: = \int^{r_1}_{r_2^n} dr \, r^2(r - \bar R)^2 
  \left[\varepsilon^n_F - V(r)\right]^{{3\over 2}} \,\,,
\eqno(2.26.2)
$$
$$
J_p: = \int^{r_1^p}_{r_2^p} dr \, r^2(r - \bar R)^2 
  \left[\varepsilon^p_F - V(r) - V_{\rm Cb}(r)\right]^{{3\over 2}} \,\,.
\eqno(2.26.3)
$$
The limits of the integrals are the (lower and
upper) classical turning points of a particle with energy
$\varepsilon_F$ in the potential $V(r)$
$$
 V(r_2^n) = V(r^n_1) = \varepsilon^n_F 
\eqno(2.27.1)
$$
$$
V(r_2^p) + V_{Cb}(r_2^p)= V(r^p_1) + V_{Cb}(r_1^p)= \varepsilon^p_F\,.
\eqno(2.27.2)
$$
We neglect the Coulomb potential in the numerical calculations.

\subsection{Spin-orbit potential}

In the conventionally used Skyrme interactions, the effective
spin-orbit interaction between two nucleons has the form
$$
 v_{\rm SO}: = i\,W_o (\vec\sigma_1 + \vec\sigma_2)\widehat{\vec k}
 \times \delta(\vec r_1 - \vec r_2) \widehat{\vec k}\,,
\eqno(2.28)
$$
where $W_o \approx 120$~MeV\,fm$^5$ and
$$
\widehat{\vec k}:  = {1\over 2i} (\vec\bigtriangledown_1 - \vec\bigtriangledown_2)\,.
\eqno(2.28')
$$
This interaction leads to the following one-body spin-orbit potential
\cite{Ring}
$$
 \widehat V_{\rm SO} = \vec W(\vec r) {1\over i} (\vec\bigtriangledown 
   \times \vec\sigma) \,\,,
\eqno(2.29)
$$
$$
\vec W(\vec r): = {3W_o\over 4} \, \vec\bigtriangledown\rho(\vec r)\,\,.
\eqno(2.29')
$$
Here $\rho(\vec r)$ is the sum of the density distribution of
neutrons ($\rho_n(\vec r)$) and protons ($\rho_p(\vec r)$).
Recently, Reinhard and Flocard \cite{Re95} generalized the ansatz
(2.28) for the spin-orbit interaction in such a~way that the
vector $\vec W(\vec r)$ (2.29') takes the more general form
$$
 \vec W_{n(p)}(\vec r) = b_4 \,\vec\bigtriangledown\rho(\vec r) + 
  b_4' \, \vec\bigtriangledown\rho_{n(p)}(\vec r)
\eqno(2.30)
$$
for neutrons and protons, resp. Here, $b_4$, $b_4'$ are
parameters in the notation of Ref. \cite{Re95}
which have the values $b_4 = b_4, = 62.130$~MeV\,fm$^5$ for the
Skyrme interaction SkI1. This choice corresponds
roughly to the value of the parameter $W_o \approx
120$~MeV\,fm$^5$, if the numbers of neutrons (N) and protons (Z)
are equal. For unequal numbers of $n$ and $p$, the form (2.30)
yields are more flexible form of the spin-orbit potential. With
the choice $b_4 = b_4' = 62.130$~MeV\,fm$^5$ one reproduces the
empirical values of the isomer shifts across magic shell
closures \cite{Re95}.

For spherically symmetric density distributions the spin-orbit
potential $\widehat V_{\rm SO}$ takes the form
$$
\widehat V^{n(p)}_{\rm SO} = {2\over r} 
  \left(b_4{\partial\rho(r)\over\partial r} + 
   b_4'{\partial\rho_{n(p)}(r) \over\partial r}\right)
  \widehat{\vec l} \cdot \widehat{\vec s}\,,
\eqno(2.31)
$$
where $\widehat{\vec s}$ is the dimesionless spin-operator and 
$\widehat{\vec l}$ is given by
$$
 \widehat{\vec l}:= \vec r \times {1\over i} \vec\bigtriangledown\,.
\eqno(2.31')
$$
For symmetric spherical nuclei
$$
 \rho_n(r) = \rho_p(r) = {1\over 2}\, \rho(r)
\eqno(2.32)
$$
one obtains the still simpler form
$$
 V^{(n)}_{\rm SO} = V^{(p)}_{\rm SO} = \widehat V_{\rm SO} = 
  {2\over r}\, (b_4 + {1\over 2}b_4') \,{\partial\rho(r)\over
  \partial r} \,\, \widehat{\vec l} \cdot \widehat{\vec s}\,,
\eqno(2.33)
$$
which is equivalent to the conventional isospin independent form
of the spin-orbit potential of a~spherical nucleus (see Eqn.
(5.103) in Ref. [11])
$$
 \widehat V_{\rm SO} = {3\over 2} \, W_o \, {1\over r} \,
  {\partial\rho(r)\over\partial r} \,\widehat{\vec l} \cdot
  \widehat{\vec s} \,\,\,.
\eqno(2.33')
$$
So far, our calculation of the spin-orbit coupling is based on
the isoscalar form (2.33') of the spin-orbit potential.
Furthermore, we used phenomenological forms for the densities
$\rho_{n(p)}(r)$ rather than the ''selfconsistent'' densities
which could be obtained from the single-particle functions. The
following density distributions were used
$$
 \rho_{n(p)}(r) = \rho^{n(p)}_o \left\{{1\over 1 +
    e^{{r-R_1\over a}}} - {1\over 1 + e^{{r-R_2\over
    a}}}\right\} 
\eqno(2.34.1)
$$
$$
 \rho_{n(p)}(r) = \rho^{n(p)}_o \left\{\left(1 + e^{{r-R_1\over a}}\right) 
    \left(1 + e^{{R_2 - r\over  a}}\right)\right\}^{-1} 
\eqno(2.34.2)
$$
$$
 \rho_{n(p)}(r) = \rho^{n(p)}_o \theta_o(r - R_2) \theta_o(R_1
   - r) (1 - \kappa(r - \bar R)^2)\,\,.
\eqno(2.34.3)
$$
where $\bar R = {1\over 2}(R_1 + R_2)$. 
The parameters $\rho^{n(p)}_o$ are determined by number
conservation 
$$
 N(Z) = \int d^3 r \,\rho_{n(p)} (r)\,.
\eqno(2.35)
$$

The results obtained from the different forms of
$\rho_{n(p)}(r)$ are very similar. The parameter $\kappa$ in
(2.34.3) is chosen in such a way that the density at $r= R_{1(2)}$
is equal $\rho^{n(p)}_0/2$ :
$$
 \kappa = {2\over (R_1 - R_2)^2} \,.
\eqno(2.36)
$$
The parabolic choice (2.34.3) of the density yields a
particularly simple form of the spin-orbit potential:
$$
 \widehat V^{n(p)}_{\rm SO} = -{12 W_o \,\rho^{n(p)}_o \over
  (R_1 - R_2)^2} \left({r - \bar R\over r}\right) \,
  \widehat{\vec l} \cdot \widehat{\vec s}\,.
\eqno(2.37)
$$
Using the unperturbed eigenstates in the coupled representation
(2.2), we obtain the shift $\Delta\varepsilon_{nlj}$ of
unperturbed eigenvalue $\buildrel\circ\over\varepsilon_{nl}$ due to
the spin-orbit potential
$$
 \Delta\varepsilon_{nlj} = {3\over 2} W_o \cdot 4 \pi
 \int^\infty_0 dr \, r \, {\partial\rho(r)\over \partial r} \,
 \buildrel\circ\over\rho_{nl}(r) \cdot \left[j(j+1) - l(l+1) - {3\over 4}
 \right]
\eqno(2.38)
$$
$$
 \varepsilon_{nlj} = \buildrel\circ\over\varepsilon_{nl} + 
    \Delta\varepsilon_{nlj}\,.
\eqno(2.39)
$$
In Eq. (2.38), the function $\buildrel\circ\over\rho_{nl}(r)$ is the
unperturbed single particle density
$$
\buildrel\circ\over\rho_{nl} = \int d\Omega \, \psi^+_{nljm}\psi_{nljm} =
 \varphi^{\star}_{nl}(r) \varphi_{nl}(r)\,.
\eqno(2.40)
$$
The total angular momentum $j$ in (2.38) can assume one of the
two values $j = |l \pm {1\over 2}|$. The approximate calculation of 
$\buildrel\circ\over\rho_{nl}(r)$ is described in Appendix A3.

\section{Results}

\subsection{Bubble nuclei in the LDM}

As we have stated already in the introduction, C.Y. Wong 
\cite{Wo73} investigated the stability of bubble
nuclei within the liquid drop (LD) model. Also the role of the 
shell effects has been first discussed in Ref. \cite{Wo73}.
The result was that
while for large enough mass ($A$) and charge numbers ($Z$)
spherical bubble configurations do have a~lower energy than the
compact spherical ones, the energy can still be lowered by
deforming the bubble. This means that no barrier protects the
spherical bubble nuclei from undergoing fission. 

It is useful to represent these results in a~simple
form. We denote the inner and outer radius of the spherical bubble
nucleus by $R_2$ and $R_1$, resp., and the radius of the compact
spherical nucleus of the same mass number by $R_0$. We retain
the conventional basic assumptions of the LDM that the charge is
uniformly distributed over the nuclear matter and that the
nuclear density is constant inside the matter distribution
dropping abruptly to zero at the surface.

The difference $\Delta E_{\rm LD}$ between the energy of the
bubble configuration with radii $R_{1,2}$ and the energy of
a~compact spherical nucleus of radius $R_0$ is given by
$$
 \Delta E_{\rm LD} = 4\pi\sigma (R^2_1 + R^2_2 - R^2_0) + 
  E_{\rm Cb}(R_1,R_2) - E_{\rm Cb}(R_0)\,,
\eqno(3.1)
$$
where $\sigma$ is the surface tension parameter.

The Coulomb energy $E_{\rm Cb}(R_1,R_2)$ of the spherical bubble
nucleus is defined as a~function of the proton density
$\rho_p(r)$ and the Coulomb potential $V_{Cb}(r)$ by
$$
E_{\rm Cb}(R_1,R_2) = {1\over 2} \int d^3r \, V_{Cb}(r) \rho_p(r)\,. 
\eqno(3.2)
$$
The LD--density $\rho_p(r)$ is represented by the simple formula
($\theta_0=$ Heaviside function)
$$ 
 \rho_p(r) = \theta_0(R_1 - r) \theta_0(r - R_2) \buildrel\circ\over\rho_p
\eqno(3.2')
$$
and the corresponding Coulomb potential is given by Eq. (2.19).
The Coulomb energy of the LD--bubble is then obtained in the form
$$
E_{\rm Cb} (R_1,R_2) = {8\pi^2 \buildrel\circ\over\rho_p^2\over 3} 
  \left[{2\over 5}
  R^5_1 + {3\over 5} R^5_2 - R^2_1 R^3_2\right]
\eqno(3.3)
$$
$$ 
 \buildrel\circ\over\rho_p = {3Ze_0 \over 4\pi R^3_0}
\eqno(3.3')
$$
The Coulomb energy $E_{\rm Cb}(R_0)$ of the homogeneously
charged sphere is 
$$
  E_{\rm Cb}(R_0) = {3Z^2e^2_0 \over 5R_0} \,\,,
\eqno(3.4)
$$
The condition of volume conservation yields the relation
$$
 R^3_0 = R^3_1 - R^3_2 \,.
\eqno(3.5)
$$
It is easily seen that the energy difference $\Delta E_{\rm LD}$
in units of the surface energy of the spherical compact nucleus
$4\pi\sigma R^2_0$ can be written as
$$
 {\cal E} :=  {\Delta E_{\rm LD} \over 4\pi\sigma R^2_0} = 
 v^2_1(v_2) + v^2_2 - 1 + 2X_0 \left[v^5_1(v_2)
 + {3\over 2} v^5_2 - {5\over 2} v^3_2 v^2_1(v_2) - 1\right]\,.
\eqno(3.6)
$$
$v_1$ and $v_2$ are the radii of the outer and inner surface in
units of $R_0$, resp.  They are related by:
$$
v_1 = {R_1 \over R_0} = \left(1 + {R^3_2\over R^3_0}\right)^{1/3} =
  (1 + v^3_2)^{1/3}
\eqno(3.7)
$$
$X_0$ is the conventional fissility parameter defined by
$$
 X_0 = {E_{\rm Cb}(R_0)\over 2\cdot 4\pi\sigma R^2_0} =
 {3Z^2e^2_0\over 40\pi\sigma R^3_0} = {(Z^2/A)\over 
 (Z^2/A)_{\rm crit}}\,.
\eqno(3.8)
$$
The dimensionless parameter $(Z^2/A)_{\rm crit}$ defined by the
last Eq. (3.8)
$$
 (Z^2/A)_{\rm crit} = {40\pi\sigma r^3_0\over 3e^2_0}
\eqno(3.8')
$$ 
depends on the value of the surface tension constant $\sigma$ and the
radius parameter $r_0$. For $r_0=1,2$~fm and
$\sigma$~=~1.026~MeV~fm$^{-2}$ one obtains $(Z^2/A)_{\rm 
crit}$~=~51.57.  We draw attention to the fact that the
difference (in units of the surface energy $E_S =
4\pi\sigma R^2_0)$ ${\cal E}$ between the energy of the bubble
configuration and the energy of the compact sphere depends only
on the fissility parameter $X_0$.

The condition of stationarity reads
$$
{\partial{\cal E} \over \partial v_2} = {2v^2_2\over v_1(v_2)} + 
  2v_2 + 2X_0 \left[-{5\over 2} v^2_2 v^2_1(v_2) + {15\over 2}
  v^4_2 - {5v^5_2\over v_1(v_2)}\right] = 0\,.
\eqno(3.9)
$$
where $v_1(v_2)$ is defined be Eq. (3.7).

Beside the trivial solution $v_2=0$ (compact sphere), the
equation (3.9) has 2 additional real solutions\footnote{Real solutions
with still larger $v_2$ may exist but they are physically meaningless.} 
if and only if $X_0 \geq$~2.02. We can write Eq. (3.9) in the form
$$
 X_0 = {2\over 5} \cdot \left[{v_2 + v_1(v_2)\over 
  v_2 + 3v^3_2 (v_2 - v_1(v_2)}\right]\,.
\eqno(3.10)
$$
The inverse $v_2(X_0)$ of (3.10) describes the values of $v_2$
which correspond to stationary values of the energy ${\cal E}$.
The function is shown in Fig. 1. From this figure, one sees in
an illustrative way that for given $X_0 > $~2.02 there are 2
real positive solutions $v_2(X_0)$. As is to be seen in Fig. 2,
the smaller one $\left(v_2^<
(X_0)\right)$ corresponds to a~maximum and the larger one
$\left(v_2^> (X_0)\right)$ to a~minimum of the energy ${\cal
E}$.

For small values of $v_2$ ($v^3_2 \ll 1$), the energy change
${\cal E}$ by bubble formation can be written approximately as
$$
{\cal E} \approx v^2_2 + {1\over 3} (2 - 5X_0) v^3_2 + 
  3X_0 v^5_2 + O(v^6_2)\,.
\eqno(3.11)
$$ 
This shows that the compact spherical LD ($v_2=0$) is always
''locally'' stable with respect to bubble formation
independently of the value of $X_0$. In other words: If there
is at all a~bubble solution, then it is separated from the
compact sphere by a~barrier.

It is a curious result that the inner and outer radius of the
bubble nucleus (measured in units of $R_0 = r_0 A^{1/3}$) depend
only on the fissility parameter $X_0$. If the parameter
$(Z^2/A)_{\rm crit}$ (see (3.8')) is independent on $(N-Z)$,
a~given value of the fissility parameter $X_0$ implies a~given
value of $Z^2/A$.

For $(Z^2/A)_{\rm crit}$=51.57, bubble
solutions within the LDM are only obtained for 
$$
 {Z^2\over A} > 2.02 \times \left({Z^2\over A}\right)_{\rm crit}
 \approx 104\,.
\eqno(3.12)
$$
Thus, for given nucleon number $A$, the charge must be larger
than a~minimal value for LD-bubble solutions to exist.
From (3.12) we obtain $Z>250$ for $A=600$ and $Z>353$ for
$A=1200$. The minimal proton density, necessary for LD-bubble
solutions to exist, is thus the larger, the smaller the nucleon
number $A$.

Let us restate clearly:{\it If we speak about the existence of
a~bubble in the LD theory, we only mean that a~minimum of the
LD-energy with respect to (volume-conserving) changes of the
bubble radii exists. As we shall show later, and as it was shown
by the author of Ref. {\rm \cite{Wo73}}, these spherical solutions
are unstable against deformation of the bubble}.

Nevertheless, it is physically meaningful to study carefully the
conditions for the existence of bubble solutions in the LDM,
because whenever these solutions exist, the shell structure effects
have a~better chance to make the bubbles stable.

The Figs. 3--5 are designed to give a global view on the
LD-properties of bubble nuclei. There results were obtained with 
the parameters of the liquid drop model taken from Ref. \cite{My66}.

In Fig. 3, lines of constant fissility parameter
$$
 X: = {E_{\rm Cb} (R_1,R_2) \over 8\pi\sigma(R^2_1 + R^2_2)}
\eqno(3.13)
$$
are shown in the $(N,Z)$-plane. Each point corresponds to an
existing bubble solution. We draw the reader's attention to the
fact that the fissility parameter $X$ defined in Eq. (3.13)
differs from the fissility parameter $X_0$, which refers to
a~compact spherical nucleus of radius $R_0$.  The two fissility
parameters are related by 
$$
 X = X_0 {v^5_1 + {3\over 2}v^5_2 - {5\over 2} v^2_1 v^3_2 \over
  v^2_1 + v^2_2} = X_0 {\left(1 + {3\over 2} f^{5/3} -
  {5\over 2} f\right) \over \left(1 - f^{5/3} - f + f^{2/3}\right)}\,,
\eqno(3.14)
$$
where the parameter $f$ is defined as the fraction 
of the empty volume to the total volume of the bubble
$$
 f = \left({R_2\over R_1}\right)^3\,.
\eqno(3.15)
$$
Contrary to $X_0$, the fissility parameter $X$ depends on the
two reduced radii of the bubble.
It is a measure of the fissility of the spherical LD bubble. The 
larger the value of $X$ the more rapid is the
descent of the energy as a~function of the deformation.

Each point on the contour lines shown in Fig. 3 corresponds to
a~bubble solution.  The parallel straight lines represent isobars. 
As one moves on a~given isobar towards
nuclei with a~larger proton density, the fissility parameter in
general decreases, i.e. the LD-bubbles become less unstable
versus fission.  A~peculiar feature is that the line of
fissility $X$=1.25 is cut twice by the isobaric lines.  This
means that for given $A$ there may exist two bubble solutions
corresponding to the same fissility $X$ but different charge
numbers and different bubble radii. Of course, the 2 solutions
belong to different binding energies in general.

Lines of equal binding energy gain $\Delta E_{\rm LD}$ (see
(3.1)) in the $(N,Z)$-plane are shown in Fig. 4.  The parallel
straight lines indicate again isobars.  On the proton rich
region of the figure for given $A$, the binding energy gains due to
bubble formation are the larger, the larger the charge $Z$.
This can be easily understood by the fact that the decrease of
the repulsive Coulomb energy due to bubble formation is larger
for larger charges.  Certain equi-energy lines are cut twice by
given isobars.  This means that there are two isobaric nuclei
$(N_1, Z_1)$ and $(N_2,Z_2)$ with bubble solutions corresponding
to the same energy gain, but in general different bubble radii.
Following the isobaric lines in the direction of larger binding
energy, one can infer the type of $\beta$-decays to be expected.
Curiously, on the $p$-rich side, $\beta^+$-decays would increase
the binding energy even further. These considerations must of
course be taken with a~grain of salt, because the bubbles are
in general not stable with regard to fission. Stability can only
be produced by shell effects which complicate the topology of
the equi-energy lines.

In Fig. 5 we display lines of equal hole parameter $f$ in the
$(N,Z)$-plane. Large $f$ means that the interior hole represents
a~large part of the total bubble volume. It is seen that in
general, as one increases the charge $Z$ along an isobaric line,
the hole fraction increases. Again there are pairs of isobaric
nuclei ($N_1,Z_1$), $(N_2,Z_2)$ with bubbles of the same shape at
different neutron and proton numbers. The energies of such pairs
are usually different.

In Fig. 6, we display lines of equal total binding energy per
particle for LD-bubbles in the $(N,Z)$-plane.  It is remarkably
different from Fig. 4 where we showed lines of constant total
binding energy gain due to bubble formation.  As a general
trend, the total LD-binding energy per particle tends to
decrease as the proton number increases along an isobaric line.
The largest binding energies per particle are seen to occur in
the region 1000$<$N$<$1800 and 300$<$Z$<$400. It is also
interesting to see an island of very small (absolute) binding
energy per nucleon in the region of 400$<$N$<$600 and
380$<$Z$<$480. Again we observe pairs of isobars showing the
same binding energy per particle.

Let us emphasize that the simple features of the Figs. 3--6 are
due to the simplicity of the LDM.  As a~macroscopic model it
predicts a~smooth dependence of all the physical properties on
$N$ and $Z$.  The shell effects, which will be shown to prevent
certain magic bubble nuclei from undergoing fission, will
complicate the picture considerably. The shell structure will be
seen to modify the smooth trends of the LDM profoundly, but only
in certain localized regions of the ($N,Z)$-plane.

So far, we only considered spherical nuclei.  A~crucial question
is whether the spherical bubbles are stable with respect to
deformations of the inner $(S_2)$ and outer $(S_1)$ surface.
Within the LD approximation, we have to study the 
energy difference
$$
 \Delta E_{\rm LD} (S_2,S_1): = E_{\rm LD}(S_2,S_1) - E_{\rm LD}(R_2,R_1)\,,
\eqno(3.16)
$$ 
where $E_{\rm LD}(S_2,S_1)$ is the energy of a deformed LD
bubble with the inner and outer surface $S_2$ and $S_1$, resp.,
and $E_{\rm LD}(R_2,R_1)$ is the LD-energy of the spherical
bubble with radii $R_2, R_1$.

As the volume terms cancel again, we have
$$
\begin{array}{ll}
\Delta E_{\rm LD}(S_2,S_1)  &= \sigma \left[S_2 + S_1 -
  4\pi R^2_2 - 4\pi R^2_1\right]  \nonumber\\
 &+ \left[{1\over 2} \int {\rho_p(\vec r_1)\rho_p(\vec r_2)\over
  |\vec r_r - \vec r_2|} d^3r_1 d^3 r_2 - E_{\rm Cb} (R_2,R_1)\right]\,,
\end{array}
\eqno(3.17)
$$
where $S_{2(1)}$ denote the magnitude of the deformed inner (outer)
surface and $\rho_p(\vec r)$ is the uniform charge distribution
within the volume enclosed by $S_2$ and $S_1$.  An investigation
of $\Delta E_{\rm LD}$ shows \cite{Wo73} that the spherical LD
bubbles are not minima but saddle points in a~multi-dimensional
potential surface. In particular, the energy of a~spherical LD
bubble is lowered by quadrupole deformations.

In Fig. 21 we show the energy $\Delta E_{\rm LD}$ as a~function
of the quadrupole deformation of the outer surface. The energy
is seen to decrease for prolate as well as for oblate
deformation of the bubble. It is also seen from this figure that
constraints on the shape of the inner surface (for given
deformation of the outer one) increase the energy substantially.

As is well-known, ordinary compact spherical liquid drops are
unstable with respect to fission for fissility parameters $X_0 >
1$. This means that the formation of a~bubble nucleus by a
fusion of 2 nuclei, is complicated by the fact that only a~tiny
part of the reaction cross section is expected to contribute to
the formation of a~bubble.

\subsection{The level spectrum}

As explained in Sect. 2, we calculated the level spectrum for
the shifted infinite square well and the shifted oscillator
treating the $\vec l\vec s$-coupling in perturbation theory. We
also studied the effects of a~finite negative constant potential in the
interior region for the case of the square well. This corresponds to 
the case of  a bubble nucleus with a reduced but nonvanishing internal 
density.

In Fig. 7, we show the level spectrum for the shifted infinite
square well as a~function of the hole fraction $f$. A~nice
feature of the shifted infinite square well is that its
eigenvalues scale with $A^{-2/3}$. We have used the unit
$100/A^{2/3}$ MeV. Consequently, from the
spectrum given in Fig. 7 one can infer the eigenvalues for
arbitrary nucleon number $A$.
We use the conventional spectroscopic notation ($nlj$):
$n=1,2,\ldots$ counts the eigenvalues for given $l,j$ in rising
order. Apart from $s,p,d\,\, (l=0,1,2)$ the angular momentum $l$ is
represented by the letters in alphabetical order ($f,g,h,\ldots
\sim l=3,4,5\ldots)$.

The following physical features of the spectrum should be noted:
for $n=1$, the single particle energies $e_{nlj}$ decrease
as a~function of $f$, for $n>2$, they increase as a~function of 
$f$ with a~steepness which becomes drastically larger with $n$.
The dependence on $n$ is produced by the requirement of a
vanishing wavefunction at the limits of the well.  As $f$
increases, the diameter $d \equiv R_1 - R_2$ diminishes and,
consequently, the energy difference between successive radial
modes increases.

The strong dependence of the difference between successive
eigenvalues of the same orbital angular momentum $l$ but
different radial modes can be understood in the following simple
approximation: If we replace the centrifugal term
${l(l+1)\hbar^2\over 2Mr^2}$ in the radial Schr\"odinger
equation by its value at the center $\bar R = {R_1 + R_2\over
2}$ of the bubble layer, the eigenfunctions $u_{nl}(r)$ become
trigonometric functions which are either ''even'' or ''odd''
with respect to the reflection $(r - \bar R) \rightarrow
-(r-\bar R)$. In this approximation the eigenenergies have the
form $(n'=0,1,\ldots)$
$$
 \varepsilon_{nl} \approx {\hbar^2l(l+1)\over 2M\bar R^2} +
 {h^2 \over 8Md^2} \cdot 
 \left\{\begin{array}{ll}
 (2n' + 1)^2 & {\rm for~even~states} \\
 (2n' + 2)^2 & {\rm for~odd~states}
\end{array}\right. \,\, .
\eqno(3.15)
$$ 
One can easily check that this relation explains qualitatively 
the behaviour of the eigenvalues as a~function of the hole
fraction $f$. We note that 
$$ 
  d^2 = R^2_0 {(1 - f^{1/3})^2 \over (1 - f)^{2/3}} 
\eqno(3.16.1)
$$
$$
 \bar R^2 = {R^2_0\over 4} {(1 + f^{1/3})^2\over (1 - f)^{2/3}} 
\eqno(3.16.2)
$$
and that the spectroscopic label $n$ is related to $n'$ by
$$
\begin{array}{llll}
 n = 1 & \sim & n' = 0, & {\rm even~state} \\
 n = 2 & \sim & n' = 0, & {\rm odd~state} \\
 n = 3 & \sim & n' = 1, & {\rm even~state} \\
   & a.s.o. & & 
\end{array}
\eqno(3.17)
$$
The gentle dependence of the energies with $n=1$ as a~function
of $f$ is due to the fact that in this case the increase of the
second term in (3.15) is largely compensated by the decrease of
the first term in (3.15).

Let us stress that the approximation (3.15) is crude and cannot
replace the determination of the true unperturbed eigenvalues by
a~numerical solution of the Eqs. (2.7).

In Fig. 7 we plotted the energy levels including the spin-orbit
interaction 
$$
e_\nu = e_{nlj} = \varepsilon_{nl} + \langle \psi_{nljm} |
 V_{\rm SO}|\psi_{nljm}\rangle\,.
\eqno(3.18)
$$
An unusual feature is that the levels with $j=l -{1\over 2}$ lie
below the levels with $j=l +{1\over 2}$. Furthermore, the
spin-orbit splitting, i.e. the energy difference $| e_{nll-{1\over 2}}
 - e_{nll+{1\over 2}}|$ is much smaller
than in the case of ordinary nuclei with the same $l$.  The
reason for both these phenomena is difference of the density
distributions: Since the spin-orbit term depends on the gradient
of the density (see Eq. (2.29'),(2.30)), its main contribution originates
from the nuclear surface. The outer surface contributes with the
same sign, but the inner with the opposite sign as compared to
ordinary nuclei. For spherical bubble nuclei the spin-orbit
potential can be written in the form (2.31). The factor
$1/r$ in the expression weighs the inner surface more
than the outer one so that the sign change becomes
understandable.  The infinite square well model is, of course,
not realistic especially for bubble nuclei with a~small value of
$f$. In the case of bubbles with a~small hole fraction, we
cannot expect the hole to be empty. It will rather be filled
with a~reduced nucleon density and thus the change at the inner
surface from the density in the layer to the reduced density in
the hole region will be much smoother than in the case of the
infinite square well. In this case, the contribution of the
outer surface may become more important than the one of the
inner surface in spite of the weight factor $1/r$ and
thus the usual order of spin-orbit splitting ensues.

In the Fig. 8 we show the level scheme as a~function of the
parameter $f$ for the shifted oscillator.  The oscillator is
more suitable than the infinite square well (i) for small
$f$-values, where the inner region is expected to contain a
reduced density of nucleons, and (ii) for bubbles with a~thin
layer, i.e.  with a~diameter $d$ which becomes smaller than
about 4~fm.  As one can infer from (3.16.1) this happens for
large values of $f$ and at the same time not too large values of
$A$. (Example: $f$=0.3 (0.4) yields $d/R_0$=0.37 (0.31)
and consequently $d$=3.9~fm (3.3~fm) for $A$=700.) Contrary to
the case of the infinite square well, there are also cases of
''normal'' order of the spin-orbit splitting (see for instance
the pairs $\left(2i{13\over 2}, 2i{11\over 2}\right)$ and
$\left(1k{17\over 2}, 1k{15\over 2}\right)$) for the harmonic
oscillator potential.

The following general trends can be stated:
\begin{itemize}
\item[(i)] For increasing hole fraction $f$, the contributions of the
inner and outer surface region to the total spin-orbit term tend
to become more equal.  Since the two regions contribute with
opposite sign, the spin-orbit splitting tends to decrease with
increasing $f$. This is seen in Fig. 7 and 8, i.e.  for both the
potentials.

\item[(ii)] The centrifugal potential has the tendency of
shifting the single particle density to larger values of $f$. For 
given hole fraction $f$, this
effect increases with $l$ and for given $l$, the effect
decreases with $f$. This shift of the s.p.  density towards
larger values of $r$ is expected to increase the contribution of
the outer surface region as compared to the inner one as far as the
spin-orbit term is concerned.
\end{itemize}

In fact one can observe in Fig. 8 that for large enough $l$ and
small $f$ the ''normal'' spin-orbit splitting prevails. The
effect is more pronounced for the oscillator than for the
square-well.

In Fig. 9 and Fig. 10 we present the level scheme for the square
well potential (2.11) with a~finite potential step $(V_2 -
V_0)~=~60$~MeV and 40~MeV, resp.  The finite internal potential
step presents the possibility to study qualitatively the effect
of a~finite reduced density inside the bubble.  In a~classical
description this was done in Ref. \cite{Mo97}. An internal density
distribution changes, of course, the surface tension of the LD
at the inner surface and, to be consistent, we should also add 
a~LD energy describing the macroscopic smooth energy originating
from the internal density. The results of Fig. 9 and 10 
are to demonstrate the effect of a~reduced internal density on the level
scheme. For this purpose we compare the level scheme in
Fig. 9 and 10 with the one in Fig. 7:

\begin{itemize}
\item[(i)] The density distribution extends from the inner surface
down to $r=0$ contrary to the case of the infinite square well
of Fig. 7. Consequently, the weight of the inner contribution to
the spin-orbit splitting increases as compared to the infinite
square well. This is seen by the fact that the splitting in the
unusual order tends to be larger in Fig. 9 and 10 as compared to
Fig. 7.

\item[(ii)] For energies far below the step, the results
are very similar to the ones of the infinite square well.

\item[(iii)] For energies above the step, the changes are
of course very large the general tendency being a~compression of
the spectrum at energies above the step.
\end{itemize}

We would like to add a~remark concerning the neutron (or proton)
numbers which appear in the plots for the 3 values: 0.12, 0.24,
0.28 of the hole fraction parameter $f$, whenever the distance
$\varepsilon_{\nu+1}-\varepsilon_\nu$ to the next higher s.p.
energy exceeds 1.5 energy units. They represent the number of
neutrons (or of protons) occupying s.p. states with energy
$\varepsilon_k \leq \varepsilon_\nu$.  Only if the total energy
of the bubble nucleus has a~minimum at one of these $f$-values,
we may expect that the numbers given represent magic numbers for
bubble nuclei. Nevertheless, they are useful to give us a~hint
as to where magic shell closures might occur. Clearly, the
precise values of the magic numbers depend on the type of shell
model used.

An interesting aspect can be observed concerning the dependence
of the ''magic numbers'' on the hole fraction variable $f$.
A general feature of all the level schemes as a~function of $f$
is the interplay of the steeply rising energies $e_{nlj}$ with
$n>1$ and the smoothly $f$-dependent energies $e_{1lj}$. Magic
numbers which appear in a~region of many crossings depend
sensitively on $f$, whereas magic numbers which occur between
$1lj$-levels away from bunches of strongly rising $2lj$- and
$3lj$-levels depend only weakly on $f$. The strongly
$f$-dependent magicity occurs preferentially for small values of
$f$, and the weakly $f$-dependent magicity for large $f$ values.

\subsection{The shell correction energy}

From the spectrum of s.p. levels $e^q_{nlj}$ we calculate the
shell correction energy in the well-known way [13] $(\nu \equiv
nlj; q = 1: neutrons, q=2: protons)$ 
$$ 
\delta E_{\rm shell} (N,Z;f) = \sum^2_{q=1}\left( \int_{-\infty}^{\lambda_q}
 e g_q(e) de \, - \, \sum_\nu e^{(q)}_\nu \right)\,\,,
\eqno(3.19)
$$
where $g_q(e)$ is the average level density:
$$
g_q(e) = {1\over \gamma\sqrt{\pi}} \sum_\nu f_{corr}(u_\nu^{(q)}) 
         \exp [-(u_\nu^{(q)})^2]
\eqno(3.20)
$$
and $f_{corr}$ is a 6th order correction polynomial of the variable
$u_\nu^{(q)}=(e-e_\nu^{(q)})/\gamma$ :
$$
f_{corr} = 1 + ({1\over 2} - u_\nu^2) + ({3\over 8} - {3\over 2} u_\nu^2
           + {1\over 2} u_\nu^4) + ({5\over 16} - {15\over 8}u_\nu^2 
           + {5\over 4} u_\nu^4 - {1\over 6} u_\nu^6)+ ....
\eqno(3.21)
$$
The position of the Fermi energy ($\lambda_q$) for the smooth energy 
spectrum is obtained by the integral:
$$
  N\,({\rm or}\,\, Z) = \int_{-\infty}^{\lambda_q} g_q(e)\,de \,\,.
\eqno(3.22)
$$
The ''plateau condition'' with respect the smearing width $\gamma$
was reasonably well fulfilled. 

In Fig. 11, we show the shell correction energy for the infinite
square well as a~function of the number of protons or neutrons.
The hole fraction parameter is chosen to be $f$=0.24. The first
substantial shell effect is seen to occur for $N$ (or $Z$)=242.
It corresponds to a~negative shell energy of about $-15\cdot\left[100 
/A^{2/3}~{\rm MeV}\right]$.  For a~doubly magic nucleus
the shell energy may amount to $(-25 {\rm \,\,to} -30) \cdot \left[100/A^{2/3}
~{\rm MeV}\right]$.  The magic numbers and the
magnitude of the shell energy depend on the hole fraction  $f$ 
(see Figs. 17--19).

In Fig. 12, the dependence of the shell energy on the height of
a~constant potential in the internal region is studied: If only
low-lying levels are occupied at a~substantially lower energy
than the step, the internal step has little influence. This is
so for particle numbers $\leq 200$.  For larger numbers of $N$
or $Z$, the magic numbers and the magnitude of the shell energy
differ noticeably from the case of the infinite square well. In
particular, the strong shell effect for 242 disappears
completely whereas a~new region of negative shell energy emerges
at 330.

For comparison we show the shell energy for a~shifted harmonic
oscillator corresponding to the same hole fraction $f=0.24$ in
Fig. 13. The frequency of the oscillator was chosen for the
particle number $A$=500 in Fig. 13.  Strong shell effects are
seen to occur for several numbers of particles of a~given sort.
They are not the same numbers as for the square well (of the
same $f$), as one would expect. Nevertheless, corresponding
shell closures are not very far away from each other and the
order of magnitude of the shell energy is the same.

The dependence of the shell energy for given ($N,Z$) on the hole
fraction $f$ is shown in the 3-dimensional plots of Fig. 14 (for
an infinite square well) and Figs. 15 and 16 (for a~harmonic
oscillator corresponding to $A$=500 and $A$=1000, resp.). It is
seen that the magnitude of the shell energy increases with $N$
(or $Z$). Each of the different islands of large negative shell
energy corresponds to specific limited intervals of the hole
fraction $f$. These intervals are of different length $\Delta l$
depending on the specific magic number. On the average, the
length $\Delta f \approx$~0.05.  The fact that magic neutron or
proton numbers for nuclear bubbles are correlated with given hole
fraction values $f$ and the fact that we expect the $n$- and
$p$-distribution in a~given bubble nucleus to exhibit the same
shape, limit the possible double magic combinations. Doubly
magic numbers for neutrons and protons can only occur in a~given
bubble nucleus if they belong to the same value of the hole
fraction $f$. This is analogous to the shell effects of ordinary
nuclei as a~function of the quadrupole deformation.

The correlation between particle number ($N$ or $Z$) and
$f$-parameter, which is visible in the Figs. 14--16, is shown
once more in the form of undistorted 2-dimensional maps in Figs.
17--19.

For ordinary exotic nuclei, especially for nuclei
with a~large excess of neutrons over protons, the distribution
of neutrons and protons do not always coincide.  In fact the
radius of the $n$-distribution is probably systematically
somewhat larger than the one of the $p$-distribution
(''$n$-skin''). For bubble nuclei with a~large $n$-excess we
have to expect the same phenomenon. Thus, for a~very $n$-rich
bubble nucleus, we expect that the inner (outer) radius of the
$n$-distribution is smaller (larger) than the inner (outer)
radius of the proton distribution. This relaxes somewhat the
condition of equal $f$-parameter for $n$ and $p$. We have
neglected this possibility in our present investigation.

Let us now discuss the total energy:
$$
 E = E_{\rm LD} + \delta E_{\rm shell}\,,
\eqno(3.23)
$$
where $E_{\rm LD}$ is the liquid drop energy with a~suitable
choice of the reference energy. So far, we studied separately
the dependence of $E_{\rm LD}$ and $\delta E_{\rm shell}$ on the
hole fraction $f$ for given $N$ and $Z$.  For this purpose we
used the LD energy of a~compact spherical nucleus (radius $R_0 =
r_0 A^{1/3}$) as a~reference energy (see Fig. 4) $$ \Delta
E_{\rm LD}: = E_{\rm LD}(f; N,Z) - E_{\rm LD}(f=0;N,Z)
\eqno(3.24)
$$ 
because in this way the volume energy cancels due to the
conventional saturation assumption.

If one adds $\delta E_{\rm shell}$ to $\Delta E_{\rm LD}$ one 
finds bubble solutions in certain regions of the $(N,Z)$-plane.
This is shown in Fig. 20.

The crucial question is the stability of these spherical bubble
configurations with respect to deformations. A~careful
investigation of the stability versus volume conserving
deformations of the inner ($S_2$) outer ($S_1$) surfaces of the
bubble nucleus requires the calculation of the surface- and
Coulomb energy of an arbitrarily shaped bubble and --- what is
much more difficult to achieve --- the calculation of the
eigenvalues of a~correspondingly deformed single particle
potential.

We have not yet performed such a~calculation. It seems
also a~bit questionable to us whether
such a~big effort is meaningful in the framework of the 
Strutinsky method, given the fact that the Hartree-Fock 
theory (HF) and Hartree-Bogoliubov theory (HB) provide
a~quantitatively more reliable approach\footnote{The existence
of stable bubble configurations for various mass-- and charge numbers
is being investigated
within the HB-theory on the basis of the Gogny-interaction
by J.~F. Berger and J.~Decharg\'e. Encouraging results
were obtained by these authors (priv. com.).}. Instead, we
applied a~simple phenomenological method which had been
introduced by W. Swiatecki \cite{My66} in 1966.   The ansatz
is designed so that the shell energy of the spherically
symmetric nucleus is damped away as a~function of the quadrupole
deformation.  
The result is shown in Fig. 21 for a~bubble nucleus of
$A$=750 and $Z$=288 and a~spherical bubble solution at $f$=0.26.
The LD-energy at the spherical bubble shape is put equal to
zero. The following features are of interest:
\begin{itemize}
\item[(i)] The LD-energy decreases for positive as well as for negative
values of the quadrupole deformation $\beta_2$ of the outer
surface $S_1$.

\item[(ii)] Constraints on the shape of the surfaces (same deformations
for the two surfaces or spherical inner surface $S_2$), 
results in noticeably higher energy than minimization of the LD-energy 
for given $\beta_2$ with respect to a limited number of shape degrees 
of $S_1$ and $S_2$.  Even spurious flat minima on the
oblate side are produced by these constraints.

\item[(iii)] $\delta E_{\rm shell} + E_{\rm LD}$ exhibit a~deep and
relatively narrow valley with a~minimum of about --30~MeV at
$\beta_2= 0$. The shell effect disappears at about $\beta\approx
\pm~0.2$ using Swiatecki's phenomenological
form of $\delta E_{\rm shell}$.

\item[(iv)] The barrier to fission is seen to be very thick. The
spontaneous fission lifetime can safely be assumed to be
practically infinite.
\end{itemize}

We can be sure that shell energies $\delta E_{\rm shell} <$ -(20 to 30)~MeV
protect the spherical bubble nuclei from undergoing fission, i.e.
lead to a~practically infinite fission lifetime.

An important question is the dependence of our results on the
parameters of the model: In the Strutinsky method, the surface
tension $\sigma$ and the radius parameter $r_0$ appear as
phenomenological quantities.  Apart from these two LD-parameters,
which enter in a significant way (see Eq. (3.8')), the form
of the phenomenological shell model and the parameters therein
influence the results. We note that the dependence of the surface tension
$\sigma$ and the radius parameter $r_0$ on $((N-Z)/A)^2$ is of great
importance for the stability regions of nuclear bubbles.  
Whereas we have some good
empirical evidence on the isospin dependence of $r_0$ \cite{Ne94}, very
little is known about the isospin dependence of $\sigma$.

In order to test how the existence of bubble solutions depends on
the parameters of the nuclear models we also performed a~calculation
based on Skyrme's energy density functional $E[\rho]$ for a~spherical
nucleus \cite{Va72}. We have assumed that the kinetic energy density
of protons $\tau_p$ and neutrons $\tau_n$ are functions of the
corresponding nucleon densities
$$
 \tau_q(r) = {\pi^{4/3}\cdot 3^{5/3}\over 5}\, \rho_q^{5/3}(r) \,\,\,\,\,
{\rm were }\,\,\,\, q=p,n\,,
\eqno(3.28)
$$
i.e. we have neglected the higher order correction which appears in the
extended Thomas Fermi approximation.
We represented the densities $\rho_n(r)$ and $\rho_p(r)$ of $n$ and $p$
by Fermi functions localized around $\bar R = {1\over 2}(R_1 + R_2)$
$$
\rho_q(r) = \buildrel\circ\over\rho_q\left({1\over 1+ exp{{r-R^q_1\over a_q}}}
          - {1\over 1+ exp{{r-R^q_2\over a_q}}}\right)
\eqno(3.29)
$$
and determined the parameters $R^q_1, R^q_2$, and the surface
thickness $a_q$ by variation of $E[\rho]$. 

Choosing the parameter set SkP proposed in Ref. \cite{Do83} and
considering a~nucleus of $A$=470 and $Z$=184 we found the
results shown in Fig. 22--25. To our surprise we obtained
a~beautiful bubble solution for a~nucleus with twice the nucleon
and proton number of $^{235}_{92}$U$_{143}$! This result is not
obtained for the more conventional Skyrme parametrization SkIII. The
specificity of the parameter set SkP is that the effective mass
of the nucleon is equal to its free mass. This fit leads to
a~more realistic level density at low excitation energy than
other Skyrme parametrizations. Here we are not pleading for
or against the interaction SkP. We only learn
from this calculation that the existence of bubble solutions for
a~given nucleus depends sensitively on the effective
interactions we use. This is to be expected given the fact that
the mass and charge of even the lightest bubble nuclei exceeds
by far those of presently known nuclei.

The results displayed in Figs. 22--25 are clearly commented in
the figure captions. We note in addition that the s.p.
potential $V(r)$ obtained (see Fig. 23) is intermediate between
a~square well and a~harmonic oscillator.  Furthermore, we see
that the energy density $h(r)$ shown in Fig. 25 drops more
smoothly to zero at the inner and outer surface than it should
for the leptodermous expansion to be valid. This demonstrates
that the LDM should not be expected to yield a~very precise value
of the bubble energy, given the fact that
the diameter $d$ of the bubble layer is in general small.

Beside the absolute minimum of the energy as a function of the hole
parameter $f$, one may find one or two competing equilibrium 
solutions at higher energy\footnote{We owe this insight to J.F. Berger
and J. Decharg\'e who found in their HFB calculations that up to
3 different bubble solutions may exist for the same number of 
neutrons and protons. For the solution with the smallest value of $f$,
these authors find the hole to be occupied by a reduced nucleon
density.}. In Fig. 26 we show results on the system with $Z$=288 and
$N$=462. The lowest solution is obtained for $f$=0.26, the 2$^{nd}$ one
with an excitation energy of 5.9 MeV at $f$=0.175, and the 3$^{rd}$
one with an excitation energy of 25 MeV at $f$=0.12 . The difference
between the total energy $E$ and the $LD$--energy in Fig. 26 represents
the shell energy. It is seen that the ground state solution is produced 
by a strongly negative shell energy in the region of $f >$0.24. 
Two other solutions are mainly produced by the appearance of a positive
shell energy in the region of 0.13$\geq f \geq$0.16 . By inspection
of the level schemes (see esp. Fig. 7) one realizes that the strongly
negative shell effect for $f >$0.24 is connected with a zone of reduced
level density for $Z$=288 below the bunch of rising levels with
1 radial node. On the other hand, in the region 0.13$\geq f \geq$0.16
the proton number 288 and the neutron number 462 correspond to Fermi
energies which are located within the bunch of rising levels with 
1 radial node.

\section{Summary and discussion}

We investigated the shell correction energy of spherical bubble
nuclei. We found strong shell effects which may give rise to
shell energies of up to -40~MeV for certain magic numbers.  By
calculating the LD-energy for deformed bubbles and assuming that
the shell effect due to the spherical symmetry disappears as
a~function of the deformation in about the same way as for
normal compact spherical nuclei, one finds that the fission
barriers are of the same order of magnitude as the
shell-correction energy for the spherical bubble solution. In
favorable cases of magic numbers, this is sufficient to reduce
the probability for spontaneous fission practically to zero.

The investigation shows that promising candidates for bubble
nuclei are found for mass numbers $A\geq 450$, with the proton
density being the larger the lighter the nucleus.

The origin of the shell effects is the high degeneracy of
orbitals with large angular momentum on the one hand, and the rapid 
energetic
increase of radial oscillation modes as a~function of increasing
bubble radius, on the other.  The details of the results (for instance the
lower limit of mass numbers, where bubble nuclei begin to exist
or the precise value of magic numbers) depend sensitively on the
choice of LD parameters and on the choice of the effective
interaction, resp.

The spin-orbit splitting for bubble nuclei is smaller than for
ordinary nuclei and the sign of the splitting may be reversed.
The reason is that the inner and outer surface regions of the
nuclear density distribution contribute to the splitting with
opposite signs (see Fig. 24).

$\beta$-decays will tend to drive the system along isobaric
lines to the composition with the largest total binding energy.
For a number of systems this will lead to desintegration by fission
after a series of $\beta$--decays. For systems where the shell
effects vary strongly as a function of $N$ and $Z$ for given $A$,
the $\beta$--decays will end up in a $\beta$--stable bubble composition.

Furthermore, a bubble nucleus consisting of $N$ neutrons and $Z$ 
protons may decay by emission of a neutron ($q$=1), a proton ($q$=2), 
or an $\alpha$--particle ($q$=3), whenever these decay channels are open,
i.e. whenever the corresponding $Q_q$--values are positive.
Semiclassically, the decay probability per time unit $W_q(Q_q,l;N,Z)$
is given by the product of the penetrability $P_q(Q_q,l)$ of the potential
barrier $U_q(r,l)$ and a quantity $\omega_q$, which can be classically
interpreted as the number of collisions per time unit of the particle
$q$ with the potential wall:
$$
W_q = \omega_q P_q \,\,,
\eqno(4.1)
$$
$$
P_q(Q_q,l) = e^{-2 J_q(Q_q,l)} \,\,,
\eqno(4.2)
$$
$$
J_q = \int_{r_1^{(q)}}^{r_2^{(q)}} d r \sqrt{{2M_q\over \hbar^2}
       \left(U_q(r;l,Z_q)-Q_q\right)} \,\,.
\eqno(4.3)
$$
Here, $M_{1(2)}$= mass of the neutron (proton) $\approx M$, $M_3$=
mass of the $\alpha$--particle, $Q_q$ are the $Q$--values of the decay
reactions, and $U_q$ represent the potential barrier between the inner 
($r_1^{(q)}$) and outer ($r_2^{(q)}$) turning points (see Fig. 27):
$$
U_q = V_q(r) + {\hbar^2 l(l+1)\over 2 M_q r^2} + {Z_q (Z-Z_q) e^2_0\over r}
\,\,.
\eqno(4.4)
$$
Here, $V_q(r)$ is the average nuclear potential felt by the particle $q$,
$Z_q$ is its charge, $Z$ is the charge of the mother nucleus, and $l$ is
the orbital angular momentum of the emitted particle. The turning points
$r_1^{(q)}, r_2^{(q)}$ are the solutions of the equation
$$
U_q(r^{(q)}_{1,2}) = Q_q \,\,.
\eqno(4.5)
$$
The frequency $\omega_q$ of assaults of the barrier may be estimated from 
the average velocity of particle $q$ in the potential well prior to emission.

The potential $V_3(r)$ which acts on the composite $\alpha$--particle
is a rough phenomenological approximation of a non--local and 
energy--dependent potential. We calculated the penetrability for an 
$\alpha$--particle through the barrier of $^{235}U$, $^{252}Fm$, and
the barrier of the bubble nucleus consisting of 470 nucleons and 182 protons
(2 times $^{235}U$). For $^{235}U$ and $^{252}Fm$ we chose experimental 
$Q$--values ($Q_\alpha (^{235}U)$=4.647 MeV and $Q_\alpha (^{252}Fm)$=8.126
MeV) and Saxon--Woods type potential $V_q(r)$ with parameters adjusted by 
Huizenga and Igo \cite{Hu62}.
For the bubble nucleus $^{470}182$ we used the $Q$-value 17.7 MeV which
was obtained from our TF--calculation based on the Skyrme--functional
and the special Skyrme interaction $SkP$ (see Figs. 22 - 25). In all the
cases the angular momentum $l$ was chosen to be 0. The barriers and
$Q$--values for 3 cases of $\alpha$--decay are shown in Fig. 27.
We obtained the following values of the $\alpha$--decay half--life
$$
T_\alpha = {ln 2\over W_3} \,\,.
\eqno(4.6)
$$
Using the experimentally known values of $T_\alpha=7\cdot 10^8 y$ for 
$^{235}U$ and $T_\alpha=25.4 h$ for $^{252}Fm$ we have obtained 
$T_\alpha=168 s$ for $^{470}182$ by means of linear extrapolation
in the ($J, log(T_\alpha)$) plane.

The system $^{470}182$ is a relatively light bubble nucleus. Note that
we obtained this bubble solution only for the special Skyrme interaction
$SkP$. As the mass and charge numbers of the bubble increases ($A >$ 750,
$Z >$ 250), the $Q_\alpha$--values tend to become larger, because the 
binding  energy per nucleon in the bubble decreases. On the other
hand, the height of the Coulomb barrier increases and the magicity of
the mother nucleus may reduce the $Q_\alpha$--value. The outcome is
that the $\alpha$--decay lifetime of heavy bubble nuclei may
become much longer than the one we obtained for $^{470}182$.

The bubble nuclei we studied were usually stable with respect to 
emission of a neutron or a proton, as one can infer from the negative slope
${d E^{LD}(N,Z)\over d N}$ and ${d E^{LD}(N,Z)\over d Z}$. In the case
of positive $Q$--values for $p$ or $n$ emission, the lifetimes can be 
estimated from the formulae (4.1)--(4.3). The decay probability $W_q$
is smaller, the larger the angular momentum $l$ of the emitted particle.
Since rather large values of $l$ occur in a typical bubble nucleus,
the additional hindrance of the decay by the centrifugal barrier may
become considerable.

Another important question is to investigate the stability of a given
bubble nucleus as a function of temperature $T$ and the angular
momentum $I$ of the system. 

We can make the following qualitative remarks: As the temperature
increases, the shell effect, which is essential for protecting
the spherical bubble from fission decreases and is known to vanish
more or less completely for $T \geq$ 1.5 MeV. As far as LD part of
the energy is concerned the main effect of a finite temperature is
reduce the surface tension. This increases the
effective fissility parameter $X_0$ (see Eq. (3.8)) and thus tends 
to favour the spherical bubble solution as compared to the energy of
the compact spherical drop (see Fig. 2). On the other hand, the
reduced stiffness of the surface implies that the LD energy of the 
bubble decreases faster as a function of deformation (see Fig. 21).
As a result, bubble nuclei cannot be expected to sustain a lot of
excitation energy.

Concerning the dependence on the angular momentum again two effects
have to be distinguished: The shell effect becomes a function
of the rotational frequency $\omega_{rot}$ of the bubble.
As a consequence, the maximal shell effect is expected to occur at
a different value of the hole parameter $f$. Furthermore, the
approximate effect of the rotation is to increase the total energy
by the rotational energy of a rigidly rotating bubble. This rotational 
energy decreases as a function of $f$. Consequently, we may expect
that an increase of the angular momentum of rotation lowers the
energy of bubble valleys as compared to the energy of a rigidly
rotating compact sphere. In specific cases, this effect may even
produce a bubble valley where there would be none for angular
momentum $I$=0 in very much the same way as this is the case for the 
2$^{nd}$ valley in the region of rare earth nuclei. However, one
should realize that the rotational energies for bubble nuclei
are the smaller the larger the mass number $A$ of the bubble nucleus.
Consequently, for heavy long--lived bubbles we may not expect a large 
effect of the rotation. Another implication of the rotation will be
that the optimal shape of the bubble will be deformed rather than 
spherical.

We have not yet taken into account pairing in bubble nuclei.
For nuclei away from shell--closures we expect the pairing energy
to be generally larger than for ordinary nuclei, because the
density of levels in the vicinity of the Fermi energy is larger
due to the high degeneracy of single particle states of large
angular momentum. The pairing energy matters only if at least one sort 
of nucleons forms a closed shell, because otherwise bubble configurations
are unstable. This means that the pairing in the bubble nucleus
is expected to play a role only for the non--magic sort of nucleons.
We do not expect that the pairing leads to a qualitatively important 
change of the stability.

It is of particular interest to investigate carefully the lower
limits of mass- and charge number where sufficiently long-lived
bubble nuclei can be expected. In view of the uncertainties
concerning our knowledge of effective interactions, these
predictions will always remain somewhat uncertain.
Nevertheless, a~lot of useful work can still be done. Assuming
that one tries to produce a~bubble nucleus of $N_b$ neutrons and
$Z_b$ protons by a~complete or incomplete fusion reaction
$$
 (N_1,Z_1) + (N_2, Z_2) = (N_b, Z_b)^\star + \cdots
\eqno(4.7)
$$
the following aspects are important:
\begin{itemize}

\item[(i)] As the compact spherical nucleus with about the same neutron and
proton numbers $N_b, Z_b$ is violently
unstable with respect to fission and other decay channels, one
may expect only a~tiny fraction of the process to lead to an
excited bubble nucleus.

\item[(ii)] Assuming that $\Delta T$ is the difference between the initial
 relative kinetic energy in the entrance channel and the energy 
$Z_1Z_2 e_0^2/(\buildrel\circ\over R_1 + \buildrel\circ\over R_2$)
of the Coulomb barrier ($\buildrel\circ\over R_i$= radius of the incident
nucleus ($N_i$, $Z_i$)), the total excitation energy $E^*$ of the bubble
nucleus can be estimated from the difference of the average binding
energy per particle $B(N_i,Z_i)$ ($<$ 0) of incident nuclei as compared 
to the average binding energy per particle $B(N_b,Z_b)$ of the bubble 
nucleus
$$
E^* = [ A_1\cdot B(N_1,Z_1) + A_2\cdot B(N_2,Z_2) - A_b\cdot B(N_b,Z_b)]
      - \Delta E + \Delta T
\eqno(4.8)
$$
Here $\Delta E$ is the energy carried away by the other final particles
in the reaction (4.7). For typical bubble nuclei, the binding energy per
particle is about -4 MeV as compared to -7 MeV for a heavy actinide
nucleus. The bracket $[ \,\, ]$ on the r.h.s. of (4.8) is then in 
general a negative number which is to be counterbalanced by the choice 
of the kinetic energy $\Delta T$. Thus the energy of the beam particles 
must be chosen considerably higher than the Coulomb barrier in order
to overcome the difference in binding energy of the constituants of
the bubble as compared to ordinary nuclei.
\end{itemize}

We draw attention to the fact that for all examples of bubble nuclei we
studied, the charge is ``overcritical'' \cite{Mu76}. This means that
the binding energy of the lowest electronic state is larger than $2m_ec^2$.
Consequently, whenever there is a K-shell vacancy in the nascent bubble, 
it will be spontaneously filled by an electron with simultaneous emission
of a positon.

\bigskip\bigskip\noindent
{\bf Acknowledgements}

\bigskip\noindent

Krzysztof Pomorski gratefully acknowledges the warm hospitality extended
to him by the Theoretical Physics Group of the Technische Universit\"at
M\"unchen as well as to the Deutsche Forschungs Gemeinshaft for granting 
a~guest professor position which enabled him to complete this research. Up 
to the 1/4/97, the BMBF supported this research. The support
made it possible to finance short term visits of K.P. at the TUM. This
support too is gratefully acknowledged.

\newpage

\section{Appendix A1}
{\large\bf Eigenfunctions of the shifted square well}
\vspace{5mm}

The eigenfunctions $\varphi_{nl}$ corresponding to the potential
(2.11) have the form
$$
 \varphi_{nl}(r) = N_{nl} \cdot
 \left\{
 \begin{array}{ll}
 f_{l}(\beta r) & {\rm for}~~0 \leq r < R_2 \\
 a_{nl} \,j_l(\alpha r) + b_{nl} \, y_l(\alpha r) & {\rm for}~~R_2 < r < R_1 
 \end{array}\right.\,,
\eqno(A1.1)
$$
where $f_l(\beta r)$ represents \cite{HMF} the spherical Bessel
function $j_l(\beta r)$ or the modified spherical Bessel
function $\tilde j_l(\beta r)$ depending on the
$sgn(\buildrel\circ\over\varepsilon_{nl} + V_1)$:
$$
 f_l(\beta r) = 
 \left\{
\begin{array}{ll}
 j_l (\beta r): = \sqrt{{\pi \over 2\beta r}} \, J_{l+{1\over 2}}(\beta r) &
   ~~~~~~~~~{\rm if}~~\buildrel\circ\over\varepsilon_{nl} + V_2 > 0\\
\tilde j_l (\beta r): = \sqrt{{\pi \over 2\beta r}} \, I_{l+{1\over 2}}
  (\beta r) &~~~~~~~~~{\rm if}~~\buildrel\circ\over\varepsilon_{nl} + V_2 < 0
\end{array}\right. \,.
\eqno(A1.2)
$$
The functions $J_{l+{1\over 2}}(\beta r)$ and
$I_{l+{1\over 2}}(\beta r)$ are the Bessel functions and the
modified Bessel functions, respectively (see Ref. \cite{HMF}, Chap. 9 and 10).
The parameter $\beta$ is defined by 
$$
 \beta = \left[{2M|\buildrel\circ\over\varepsilon_{nl} + V_2|\over\hbar^2} 
 \right]^{1/2}\,.
\eqno(A1.3)
$$
The dimensionless amplitudes $a_{nl}$, $b_{nl}$ and the
eigenenergies $\buildrel\circ\over\varepsilon_{nl}$ are determined by
the boundary and continuity requirements
$$
 f_l(\beta R_2) =a_{nl} \, j_l(\alpha R_2) + b_{nl} \, y_l(\alpha R_2)
\eqno(A1.4)
$$
$$
  \beta f'_l(\beta R_2)
 =  \alpha\left[a_{nl} \, j'_l(\alpha R_2) + b_{nl} \, y'_l(\alpha R_2)\right]
\eqno(A1.5)
$$
$$
0 = a_{nl} \, j_l(\alpha R_1) + b_{nl} \, y_l(\alpha R_1)\,.
\eqno(A1.6)
$$
Here and in what follows we use the notation:
$$
j'_l(\alpha R_2) \equiv \left({\partial j_l(z)\over \partial
  z}\right)_{z=\alpha R_2} ~~~~~
\mbox{and equally for}~~~y'_l(\alpha R_2), f'_l(\beta R_2)\,.
$$
We satisfy the eqns. (A1.6) by the ansatz
$$
 a_{nl} = \tilde a_{nl} \, y_l(\alpha R_1)
\eqno(A1.7)
$$
$$
 b_{nl} = - \tilde a_{nl} \, j_l(\alpha R_1)
\eqno(A1.8)
$$
The amplitude $\tilde a_{nl}$ is then obtained from (A1.4)
$$
\tilde a_{nl} = {f_l(\beta R_2)  \over
  y_l(\alpha R_1) j_l(\alpha R_2) - j_l(\alpha R_1) y_l(\alpha R_2)}\,.
\eqno(A1.9)
$$
and the eigenenergies $\buildrel\circ\over\varepsilon_{nl}$ from 
the continuity of the logarithmic derivatives
$$
\beta {f'_l(\beta R_2) \over f_l(\beta R_2)} =  \alpha \,
 {y_l(\alpha R_1) j'_l(\alpha R_2) - j_l(\alpha R_1)
   y'_l(\alpha R_2) \over y_l(\alpha R_1) j_l(\alpha R_2) -
   j_l(\alpha R_1) y_l(\alpha R_2)}\,.
\eqno(A1.10)
$$
The overall normalization constant $N_{nl}$ is of course defined by the condition
(2.10).

\section{Appendix A2}
{\large\bf Eigenfunctions of the shifted harmonic oscillator}
\vspace{5mm}

For $r < R_2$, the radial wave functions are again given by
(A1.1) and (A1.2) up to a factor $c_{nl}$ (see (A2.4)). At the point 
$r = R_2$, the radial wave
function $\varphi_{nl}(r)$ in region I ($\varphi^{\rm
I}_{nl}(r)$) and in region II ($\varphi^{\rm II}_{nl}(r)$) must be continuous and also its derivatives:
$$
 \varphi^{\rm I}_{nl}(R_2) = \varphi^{\rm II}_{nl}(R_2)
\eqno(A2.1)
$$
$$
\frac{d\varphi^{\rm I}_{nl}(R_2)}{dR_2} = 
\frac{d\varphi^{\rm II}_{nl}(R_2)}{dR_2}  \,.
\eqno(A2.2)
$$
Furthermore, $\varphi^{\rm I}_{nl}(r)$ must satisfy
(2.5.2) at infinity.

The eigenvalues $\buildrel\circ\over\varepsilon_{nl}$ and the wave functions
$\varphi^{\rm I}_{nl}(r)$ must be determined
numerically or by the WKB approximation. In both cases it is
convenient to introduce the radial functions 
$u^I_{nl}(\eta)$ depending on the
dimensionless variable $\eta$
$$
\eta = \gamma r = \sqrt{{M\omega\over\hbar}} \cdot r
\eqno(A2.3)
$$
for describing the functions $\varphi_{nl}(r)$ in region I:
$$
\varphi_{nl}(r) =  N_{nl}
\left\{
 \begin{array}{ll}
 c_{nl} \, f_l(\beta r) & {\rm for}~~0 \leq r < R_2 \\
   {u^I_{nl}(\eta)\over \eta} & {\rm for}~~r > R_2 \,\,\,\,\,.
\end{array}\right.
\eqno(A2.4)
$$
Henceforth, we shall suppress the subscripts $nl$ for simplicity.
The radial function $u^{\rm I}(\eta)$ satisfies the differential
equation 
$$
{d^2 u^{\rm I}(\eta)\over d\eta^2} - \left[{l(l+1)\over \eta^2}
+ (\eta - \bar\eta)^2 + \kappa\right] \, u^{\rm I}(\eta) = 0\,,
\eqno(A2.5)
$$
where 
$$
\bar\eta = \gamma\bar R
\eqno(A2.6)
$$
$$
 \kappa : = {2(\buildrel\circ\over\varepsilon + V_0) \over 
 \hbar\omega}\,.
\eqno(A2.7)
$$
For large values of $\eta$, the centrifugal term ${l(l+1)\over
\eta^2}$ in (A2.5) is negligible and the equation (A2.5)
assumes the standard form of the differential equation for
parabolic cylinder functions (see Ref. \cite{HMF}, p. 686):
$$
 {d^2 w \over d\xi^2} - {1\over 4}\xi^2 w(\xi) + {\kappa\over 2} \,
  w(\xi) = 0\,,
\eqno(A2.8)
$$
when introducing the new independent variable, 
$$
\xi : = \sqrt{2}(\eta - \bar\eta)\,.
\eqno(A2.9)
$$
The functions $u^{\rm I} (\eta)$ approach the solutions $w(\xi)$
of Eq. (A2.8)
$$
 u^{\rm I}(\eta) \rightarrow w(\kappa;\xi) = w[\kappa; \sqrt{2}(\eta -\bar\eta)]
\eqno(A2.10)
$$
for $\eta$-values satisfying the inequality
$$
 \eta^2(\eta - \bar\eta)^2 \gg l(l + 1)\,.
\eqno(A2.11)
$$
For obtaining normalizable eigenfunctions $\varphi_{nl}(r)$ we
need solutions $w(\xi)$ of the differential equation (A2.8)
which go to zero at infinity ($\xi \rightarrow\infty$) at least
$\sim\xi^{-1}$. The mathematical properties of the two
basic, linearly independent solutions $y _{1,2}$ of Eq. (A2.8)
(see Eq. (A2.13) and (A2.14) are described in Chapter 19 of
Ref. \cite{HMF}. The special linear combination which vanishes
for $\xi \rightarrow \infty$ is given by
$$
 w(\kappa;\xi) = \sqrt{\pi} 2^{{\kappa\over 4}} \left[{y_1(\xi) 
 \over 2^{1\over 4}\Gamma\left({3-\kappa\over 4}\right)} -
 {2^{{1\over 4}}y_2(\xi) \over
 \Gamma\left({1-\kappa\over 4}\right)}\right] 
\eqno(A2.12)
$$
$$
y_1(\xi) = e^{-{\xi^2\over 4}} \, M\left({1-\kappa\over 4},
 {1\over 2}; {\xi^2\over 4}\right)
\eqno(A2.13)
$$
$$
y_2(\xi) = \xi \, e^{-{\xi^2\over 4}} \, M\left({3-\kappa\over 4},
 {3\over 2}; {\xi^2\over 4}\right)\,.
\eqno(A2.14)
$$
Here, $M(a,b;z)$ is the confluent hypergeometric function
$$
 M(a,b;z) = 1 + {a\over b} \, {z\over 1!} + 
 {a(a+1)\over b(b+1)}  \, {z^2\over 2!} + \cdots
\eqno(A2.15)
$$
It is interesting to consider the special case of $\kappa = 2n +
1$ (where $n=0,1,2,\ldots$), where the functions $w(\kappa;\xi)$
become equal to the even Hermite functions for even $n$ and to
the odd Hermite functions for odd $n$ (see Ref. \cite{HMF}, Eq.
(19.13.1))
$$
 w(\kappa = 2n+1; \xi) = e^{-{1\over 4}\xi^2} \, He_n(\xi)
\eqno(A2.16)
$$
$$
He_n(\xi) : = (-1)^n \, e^{{\xi^2\over 2}} \, {d^n\over d\xi^n}
 \left\{e^{-{\xi^2\over 2}}\right\}\,.
\eqno(A2.17)
$$
i.e. to the eigenfunctions of the linear harmonic oscillator. Of
course, the functions $u^{\rm I}(\eta)$ do not even
asymptotically tend to this limit because of the boundary
conditions (A2.1-2) at $r = R_2$ and because of the centrifugal
term in (A2.5). In the case of a~general value of $\kappa$, the
functions $w(\kappa;\xi)$ represent ''parabolic cylinder
functions'' and are frequently denoted by (Ref. \cite{HMF}, Eq.
(19.3.1)
$$
 w(\kappa;\xi) = {\cal D}_{{1\over 2}(\kappa-1)} (\xi)\,.
\eqno(A2.18)
$$
For $|\xi| \gg {|\kappa|\over 2}$, they have the following
asymptotic form (Ref. \cite{HMF}, (19.8.1)):
$$
 w(\kappa;\xi) \approx e^{-{1\over 4}\xi^2} \, 
  \xi^{{\kappa-1\over 2}} \left\{1 - {(\kappa-1)(\kappa-3) 
  \over 2^3\xi^2} + {(\kappa-1)(\kappa-3)(\kappa-5)(\kappa-7) 
  \over 2^5 \cdot 4\xi^2} - \cdots \right\}\,. 
\eqno(A2.19)
$$
Solutions $u^{\rm I}(\eta)$ which tend asymptotically to the
parabolic cylinder function $w(\kappa;\xi)$ (see (A2.10)) are
thus normalizable for all values of $\kappa$.

The precise values of $\kappa$ (or $\varepsilon_{nl}$) and of the
amplitude $a_{nl}$ are determined by the boundary conditions
(A2.1-2). Substituting (A2.4) these boundary conditions read
$$
 c_{nl} \, f_l(\beta R_2) = {u^{\rm I}_{nl}(\gamma_2 R_2)\over 
 \gamma R_2} 
\eqno(A2.20)
$$
$$
 c_{nl}\,\beta \left({df_l(z)\over dz}\right)_{z=\beta R_2} =
 \gamma \cdot \left({{du^{\rm I}_{nl}(\eta)\over d\eta} -
  u^{\rm I}_{nl}(\eta)\over \eta^2}\right)_{\eta=\gamma R_2}\,.
\eqno(A2.21)
$$
The functions $u^{\rm I}_{nl}(\eta)$ can be determined
numerically by the ''shooting method''.

\section{Appendix A3}
{\large\bf Thomas-Fermi approximation for eigenenergies
$\varepsilon_{nl}$ and single-particle densities $\rho_{nl}(r)$.} 
\vspace{5mm}

We mentioned already the WKB-approximation for the unperturbed
eigenvalues $\varepsilon_{nl}$ in Chapter 2. The Eq.
(2.15) represents the simplest version of the WKB-approximation
which contains no correction from the vicinity of the turning
points nor from the classically forbidden regions. In what
follows we only refer to this simplest version and claim that it
can essentially also be obtained from the Thomas-Fermi
approximation (TFA).  

Let us first assume that a given eigenvalue $\varepsilon_{nl}$
 has been calculated from the simple
semiclassical quantization condition (2.15) which we now write
in the form
$$
\sqrt{2M} \int^b_a dr \sqrt{\buildrel\circ\over\varepsilon_{nl} - U_l(r)}
= \left(n + {1\over 2}\right) \hbar\pi\,,
\eqno(A3.1)
$$
where the potential $U_l(r)$ is defined by
$$
 U_l(r) = \left\{
\begin{array}{ll}
 V(r) + {l(l+1)\hbar^2\over 2Mr^2} & {\rm for~ neutrons} \\
V(r) + e_0 V_{Cb}(r) + {l(l+1)\hbar^2\over 2Mr^2} & {\rm for~ protons} 
\end{array}\right.
\eqno(A3.2)
$$
and where $n=0,1,2,\ldots$.

As we stated already in Sect. (2.1), the Coulomb potential
$V_{\rm Cb}(\vec r)$ produced by the density distribution
$\rho_p(\vec r)$ of the protons (see Eq. (2.16)) can be easily
evaluated in closed form if we approximate the density
$\rho_p(\vec r)$ by a~uniform distribution in the bubble layer
(see (2.18)). The result is given in Eq. (2.19).

The spin-orbit potential $\widehat V_{\rm SO}$ ((2.33') or
(2.37)) acting on eigenstates (2.2) in the coupled
representation is a $c$-number operator.  Consequently, in order
to evaluate its mean value, we only need the single particle
densities $\buildrel\circ\over\rho_{nl}(r)$ as defined in Eq. (2.40).
Since no derivatives occur, the accuracy requirements
concerning these single-particle densities are less
stringent.  Therefore, we determined these densities in the
Thomas-Fermi approximation (TFA).

The numbers $N$ of neutrons and $Z$ of protons are related to
the Fermi energies $\varepsilon^n_F$ and $\varepsilon^p_F$ by
the Eqs. (2.25). On the
other hand, the total densities $\buildrel\circ\over\rho_n(r)$, 
$\buildrel\circ\over\rho_p(r)$ of
$n,p$ must satisfy the trivial relations 
$$
 N = \int d^3 r \,\buildrel\circ\over\rho_n(r)\,\,,
\eqno(A3.3)
$$
$$
 Z = \int d^3 r \,\buildrel\circ\over\rho_p (r)\,\,.
\eqno(A3.4)
$$
By comparison with (2.25) we obtain the TFA of the densities:
$$
\buildrel\circ\over\rho_n(r) = {2\over h^3} \int d^3p \, \theta_0 
  \left[\varepsilon^n_F - {p^2\over 2M} - V(r)\right] =
 {8\pi(2M)^{3/2}\over 3 h^3} [\varepsilon^n_F - V(r)]^{3/2}
\eqno(A3.5)
$$
$$
\buildrel\circ\over\rho_p(r) = {2\over h^3} \int d^3p \, \theta_0 
  \left[\varepsilon^p_F - {p^2\over 2M} - V(r) - V_{\rm Cb}(r)\right] =
{8\pi(2M)^{3/2}\over 3 h^3} [\varepsilon^p_F - V(r) - V_{\rm Cb}(r)]^{3/2}\,.
\eqno(A3.6)
$$ 
By integrating over the phase space enclosed by two
neighboring surfaces corresponding to the constant energies
$\left(\varepsilon_\nu - {\Delta\varepsilon\over 2}\right)$ and
$\left(\varepsilon_\nu + {\Delta\varepsilon\over 2}\right)$ we
obtain the number of $n$ or $p$ contained in this phase space.
Requiring that the volume is occupied by 1 $n$ or $p$ we may
determine the corresponding energy width $\Delta\varepsilon$
which depends on the energy $\varepsilon_\nu$ and may be
insignificantly different for $n$ and $p$: 
$$ 
1 = {1\over h^3} \int d^3r  \, \int d^3p\, \theta_0 
\left(\varepsilon_\nu + {\Delta\varepsilon\over 2} - {p^2\over 2M}
 - \tilde V(r)\right) \theta_0\left({p^2\over 2M} + 
 \tilde V(r) - \varepsilon_\nu + {\Delta\varepsilon\over 2}\right)\,.
\eqno(A3.7)
$$
Here $\tilde V(r)$ is the potential acting on a $n$ or $p$, resp.:
$$
\tilde V(r) = V(r)~~~~~~~~~~~~{\rm for~neutrons} 
\eqno(A3.8)
$$
$$
\tilde V(r) = V(r) + e_0 V_{\rm Cb}(r)~~{\rm for~protons} \,.
\eqno(A3.9)
$$
The density distribution $\buildrel\circ\over\rho(r;\varepsilon_\nu)$ 
of the neutrons (protons) in the energy layer is given by
$$
\buildrel\circ\over\rho(r;\varepsilon_\nu) = {2\over h^3} \int d^3p \,
 \theta_0 
  \left(\varepsilon_\nu + {\Delta\varepsilon\over 2} - {p^2\over 2M}
 - \tilde V(r)\right) \theta_0\left({p^2\over 2M} + 
 \tilde V(r) - \varepsilon_\nu + {\Delta\varepsilon\over 2}\right)\,.
\eqno(A3.10)
$$
or
$$
\buildrel\circ\over\rho(r;\varepsilon_\nu) = {8\pi\over 3h^3} (2M)^{3/2}
  \left[\left(\varepsilon_\nu + {\Delta\varepsilon\over 2} -
   \tilde V(r)\right)^{3/2} - \left(\varepsilon_\nu -
   {\Delta\varepsilon\over 2} -  \tilde V(r)\right)^{3/2}\right]\,,
\eqno(A3.11)
$$
where the width $\Delta\varepsilon$ of the energy layer is to be 
determined from relation (A3.7).
The condition (A3.7) reads more explicitly
$$
1 = {(4\pi)^2\over 3h^3}(2M)^{3/2} \, \int dr \, r^2
  \left[\left(\varepsilon_\nu + {\Delta\varepsilon\over 2} -
   \tilde V(r)\right)^{3/2} - \left(\varepsilon_\nu -
   {\Delta\varepsilon\over 2} -  \tilde V(r)\right)^{3/2}\right]\,. 
\eqno(A3.12)
$$
In each case, integrations over the radial coordinate $r$
extend over the interval where the integrand is real. 

For calculating the average of the spin-orbit potential
$\widehat V_{\rm SO}$ we need the single particle density
$\buildrel\circ\over\rho_{nl}(r)$ of a~nucleon of given energy 
$\varepsilon_{nl}$ and given orbital angular momentum $l$.
The simple form (A3.11) of the density cannot be used because
it averages over particles of all angular momenta $l$.
From the TFA we cannot obtain a~single particle density with the desired
number $n$ of radial nodes but we may obtain an
approximate density distribution for a~nucleon of given orbital
angular momentum $l$ by performing the integration in (A3.7) and
(A3.10) under the constraint
$$
\vec L^2 \equiv (\vec r \times \vec p)^2 = l(l + 1)\hbar^2\,.
\eqno(A3.12)
$$
This can be simply achieved by choosing the 3-axis of the
momentum integration in the direction of the position vector
$\vec r$ and using cylinder coordinates in $p$-space:
$$
 \vec p = p_3 \, \vec e_r + p_\perp (\cos\phi_p \, 
  \vec e_1 + \sin\phi_p \, \vec e_2)\,,
$$
$$
 d^3p = dp_3 \,dp_\perp p_\perp d\phi_p\,,
$$
$$
\vec L^2 = L^2 = r^2 p^2_\perp
$$
$$
{\vec p}^2 = p^2_3 + p^2_\perp = p^2_3 + {L^2\over r^2} \,.
$$
The condition (A3.12) can be fulfilled conveniently by introducing
the continuous dimensionless integration variable $\lambda$
instead of $p_\perp$
$$
p_\perp = {L\over r} = {\lambda\hbar \over r}
$$
and inserting the $\delta$-function $\delta\left(\lambda -
\sqrt{l(l+1)}\right)$ into the integrals. We thus obtain the following
2 relations in analogy to the Eqs. (A3.7) and (A3.10)
$$
\begin{array}{rcl}
1 & = & {\hbar^2\over h^3} \int d^3r \int dp_3 \, d\lambda \,
  d\phi_p \, {\lambda\over r^2} \,\theta_0   
  \left(\varepsilon_{nl} + {\Delta\varepsilon_l\over 2} - {p^2_3\over 2M}
 -  {\lambda^2\hbar^2 \over 2Mr^2} - \tilde V(r)\right)  \\
& \cdot &   \theta_0 \left({p^2_3\over 2M} + {\lambda^2\hbar^2 \over 2Mr^2}
  + \tilde V(r) - \varepsilon_{nl} + {\Delta\varepsilon_l\over
   2}\right)\, \delta \left(\lambda - \sqrt{l(l+1)}\right) \\
\end{array}
\eqno(A3.13)
$$
$$
\begin{array}{rcl}
\buildrel\circ\over\rho_{nl}(r;\varepsilon_{nl}) & = 
  &{\hbar^2\over h^3r^2} \int dp_3\, d\lambda \,
  d\phi_p \,\, \lambda\, \theta_0 \left(\varepsilon_{nl} +
  {\Delta\varepsilon_l\over 2} - {p^2_3\over 2M} 
  -  {\lambda^2\hbar^2 \over 2Mr^2} - \tilde V(r)\right)  \\
& \cdot &  \theta_0 \left({p^2_3\over 2M} + {\lambda^2\hbar^2 \over 2Mr^2}
  + \tilde V(r) - \varepsilon_{nl} + {\Delta\varepsilon_l\over  2}
   \right) \delta\left(\lambda - \sqrt{l(l+1)}\right)\,. 
\end{array}
\eqno(A3.14)
$$
Carrying out the integrations in (A3.14) we obtain 
$$
\buildrel\circ\over\rho_{nl}(r; \varepsilon_{nl}) = 
 {2\sqrt{2M}\sqrt{l(l+1)}\over 2\pi h r^2}
\left\{\sqrt{\varepsilon_{nl} + {\Delta\varepsilon_l\over 2} - U_l(r)} - 
  \sqrt{\varepsilon_{nl} - {\Delta\varepsilon_l\over 2} - U_l(r)}\right\}\,.
\eqno(A3.15)
$$
where $U_l(r)$ represents the potential including the
centrifugal term (see Eq. (A3.2)). 
For $l\gg 1$, we have $2\sqrt{l(l+1)} \approx 2l$.
This represents the classical approximation for the
quantum-mechanical degeneracy factor ($(2l+1)$. The relation
(A3.13) takes the form:
$$
1 = {2\sqrt{l(l+1)}2\sqrt{2M}\over h} \int dr
  \left\{\sqrt{\varepsilon_{nl} + {\Delta\varepsilon_l\over 2} - U_l(r)} - 
  \sqrt{\varepsilon_{nl} - {\Delta\varepsilon_l\over 2} - U_l(r)}\right\}\,.
\eqno(A3.16)
$$

The main deficiency of the TFA is to completely neglect the
density in the classically forbidden region and to ignore the
correct nodal structure of the density distribution of single particles
inside the potential well.

The TFA is closely related to the WKB approximation. Let us
consider the Hamiltonian
$$
 \widehat H_0 = -{\hbar^2\over 2M} {d^2\over dx^2} + V(x)
\eqno(A3.17)
$$
of a 1-dimensional potential model and the corresponding
classical Hamiltonian function
$$
 {\cal H} = {p^2 \over 2M} + V(x)\,.
\eqno(A3.17')
$$
If the particles moving in the potential $V(x)$ are fermions of
spin ${1\over 2}$, the number ${\cal N}(\varepsilon_\nu)$ of
fermions which fill the energy levels below a~given energy
$\varepsilon_\nu$ are given in the TFA by
$$
 {\cal N}(\varepsilon_\nu) = {2\over h} \int dx \, \int dp \, 
 \theta_0 \left(\varepsilon_\nu - {\cal H}(x,p)\right)\,.
\eqno(A3.18)
$$
We can use the relation (A3.18) in order to determine approximate
eigenvalues $\varepsilon_1 < \varepsilon_2 < \ldots$ by
requiring ${\cal N}(\varepsilon_\nu)$ to be equal to $2\nu$, where
$\nu =1,2,\ldots$ counts the eigenvalues in rising order. We thus
obtain the implicit equation for determining the eigenvalues
$$
\nu = {1\over h} \int dx \, \int dp \, \theta_0 \left(\varepsilon_\nu
    - {\cal H}(x,p)\right)\,. 
\eqno(A3.19)
$$
For the linear harmonic oscillator $V(x) = {M\omega^2\over 2}
x^2$ we find from (A3.19) the eigenvalues
$$
 \varepsilon_\nu = \nu\hbar\omega
\eqno(A3.20)
$$
for $\nu = 1,2,\ldots$. Using the same definition of the index
$\nu$, the correct eigenvalues are given by
$$
 \varepsilon_\nu = (\nu - {1\over 2})\hbar\omega\,.
\eqno(A3.21)
$$
For the 1--dimensional infinite square well, the TFA (A3.19) yields
the correct result.
In the case of a spherical potential in 3-dimensional space we have
to restrict the integration of the phase space to the hypersurface 
defined by the constraint (A3.12), i.e. we apply the method for
calculation of the density distribution $\buildrel\circ\over\rho_{nl}$. 
This yields
$$ 
 n \cdot (2l+ 1) = {\hbar^2 \cdot 2\pi\sqrt{l(l+1)} \over h^3}
 \int d^3 r \int dp_3 \, \theta   \left(\varepsilon_{nl} 
 - {p^2_3\over 2M} - U_l(r)\right) {1\over r^2}\,,
\eqno(A3.22)
$$
where $n=1,2,\ldots$ and $(2l+1)$ is the degeneracy of a level
of given $l$. Within our approximation we should put
$2\sqrt{l(l+1)} \approx 2l \approx 2l+1$. Consequently, (A3.22)
takes the form 
$$
 \sqrt{2M} \int^b_a dr \sqrt{\varepsilon_{nl} - U_l(r)} = 
  n\hbar\pi\,\,,
\eqno(A3.23)
$$
where $a$ and $b$ are the lower and upper turning points
and $n=1,2,\ldots$. The Eq. (A3.23) resembles the WKB formula
given by Eq. (2.15). We note, however, that in (A3.23)
the radial quantum number $n$ starts from 1 whereas in (2.15)
it starts from 0. For $n\gg 1$, the results are equivalent.

\newpage

{\bf Figures captions.}

\begin{enumerate}
\item[ 1.] The function $v_2(X_0)$ defining the stationary points of the 
           energy ${\cal E}(v_2;X_0)$ as a function of the fissility 
           parameter $X_0$.

\item[ 2.] Difference ${\cal E}= \Delta E_{\rm LD}/E_S(R_0)$ of the LD energy of 
           a bubble of reduced inner radius $v_2$ and the LD energy of 
           a compact spherical nucleus of radius $R_0$. 

\item[ 3.] Lines of equal bubble fissility parameter $X$ in the $(N,Z)$ 
           plane. Isobars are indicated by the parallel straight lines.

\item[ 4.] Lines of equal binding energy gain $\Delta E_{\rm LD}$ by bubble
           formation. Parallel straight lines indicate isobars.

\item[ 5.] Lines of equal hole fraction $f$ in the $(N,Z)$ plane.
           Parallel straight lines indicate isobars.

\item[ 6.] Lines of constant LD--binding energy per particle. 
           Parallel straight lines indicate isobars.

\item[ 7.] Level scheme for the shifted infinite square well with the 
           spin--orbit coupling  as a function of the hole fraction 
           $f=(R_2/R_1)^3$ (see Eq. (3.15)). The letters s, p, d, f, g, h,
           i, j, ... mean the orbital angular momenta
           $l = 0, 1, ...$ . For 3 values of $f$ (0.12, 0.24, 0.28) numbers 
           are written just above certain single particle energies 
           $\varepsilon_\nu$ whenever the distance
           ($\varepsilon_{\nu +1} - \varepsilon_\nu$) exceeds 1.5 energy 
           units. They represent the number of neutrons (or of protons)
           occupying single particle states with energy $\varepsilon_k \leq
           \varepsilon_\nu$.
      
\item[ 8.] Level scheme for the shifted harmonic oscillator with spin--orbit
           coupling as a function of the hole fraction $f$. The 
           oscillator frequency $\omega$ was determined as a function of 
           $f$ for $A$=500.

\item[ 9.] Level scheme for the infinite square well potential with a~finite 
           step in the inner region. Height $(V_0-V_2)$ of the step: 60 MeV 
           for $r<R_2$.

\item[10.] Same as in Fig. 9, but with a step height of 40 MeV.

\item[11.] Shell correction energy $\delta E_{shell}$ for bubbles with
           $f$=0.24 as a function of $N$ or $Z$ for the shifted infinite 
           square well with spin--orbit term.

\item[12.] Same as in Fig. 11, but for a square well with an internal step 
           height $(V_0-V_2)$= 
           40, 50, and 60 MeV. The case of the simple infinite square well
           is given again for comparison.

\item[13.] Same as in Fig. 11, but for a shifted harmonic oscillator 
           potential. The oscillator frequency corresponds to the mass number
           $A$=500. 

\item[14.] Shell energy evaluated from the infinite square well
           plus $\vec l\cdot\vec s$--term as a function of $N$ (or $Z$) 
           and the hole fraction $f$. The energy units are [$100/A^{2/3}$]
           MeV.

\item[15.] As in Fig. 14, but for a shifted harmonic oscillator 
           potential corresponding to the particle number $A$=500.
           Energy unit: MeV.

\item[16.] As in Fig. 14, but for a shifted harmonic oscillator 
           potential corresponding to the particle number $A$=1000.
           Energy unit: MeV.

\item[17.] Lines of constant shell energy as a function of particle number
           ($N$ or $Z$) and the hole fraction parameter $f$ for a shifted
           infinite square plus $\vec l\cdot\vec s$--coupling. The energy
           unit is [$100/A^{2/3}$] MeV.

\item[18.] As in Fig. 17, but for a shifted harmonic oscillator with 
           $\vec l\cdot\vec s$--coupling. The oscillator frequency
           corresponds to the particle number $A$=500.
           Energy unit: MeV.

\item[19.] As in Fig. 18, but with the oscillator frequency chosen for 
           the mass number $A$=1000 .

\item[20.] The map of the liquid drop plus the shell correction energy
	   corresponding to the bubble solution.

\item[21.] LD--energy (solid line), shell correction energy (dotted line), 
           and the total energy (dashed dotted line) as a function of the 
           quadrupole deformation $\beta_2$ of the outer bubble surface 
           $S_1$. The LD--energy is minimized with respect to the multipole
           deformation of the order 4 and 6 of the surface $S_1$ and the
           multipole 
           deformation 2 to 6 of the inner surface $S_2$. The long dashed
           line represents the LD-energy when both surfaces have the same
           deformation. The short dashed line corresponds to the case when
           the inner surface is a sphere.

\item[22.] Density $\rho_p(r)$ of protons (solid line) and $\rho_n(r)$ of
           neutrons (dashed line) obtained by variation of the Skyrme
           functional (3.28) obtained within the EFT approximation for SkP 
           forces and for the nucleus Z=184, A=470.

\item[23.] Nuclear central s.p. potential acting on protons (solid line)
           and neutrons (dashed line) as obtained from the variation of the
	   Skyrme energy functional \cite{Va72}.

\item[24.] Average spin--orbit potential $V_{LS}(r)$ for $A$=470 and $Z$=184.

\item[25.] Energy density $h(r)$ for $A$=470 and $Z$=184.

\item[26.] Total energy $E$ and liquid drop energy $E_{LD}$ as a function
           of the hole parameter $f$ for a system with $Z$=288 protons
           and $A$=750 nucleons. 3 bubble solutions corresponding to the
           $f$-values $f$=0.26 (ground state), $f$=0.175 (bubble valley
           at the excitation energy $E^*$=5.9 MeV), and $f$=0.12 (bubble
           valley at the excitation energy $E^*$=25 MeV) are seen to exist.
           The inner and outer radii of the 3 solutions are:
           $R_2$=7.69 fm and $R_1$=12.05 fm for $f$=0.26,
           $R_2$=6.50 fm and $R_1$=11.62 fm for $f$=0.175,
           $R_2$=5.61 fm and $R_1$=11.37 fm for $f$=0.12 .

\item[27.] The potential barrier $U_3(r)$ acting on an $\alpha$--particle
           in $^{235}U$, in $^{252}Fm$, and in the bubble nucleus $^{470}184$
           are shown together with the corresponding $Q_3=Q_\alpha$ values.
           The bubble nucleus $^{470}184$ is the one obtained as a result of 
           a TF--calculation with the Skyrme interaction $SkP$ (see Figs.
           22 - 25).
\end{enumerate}


\newpage
\ifig{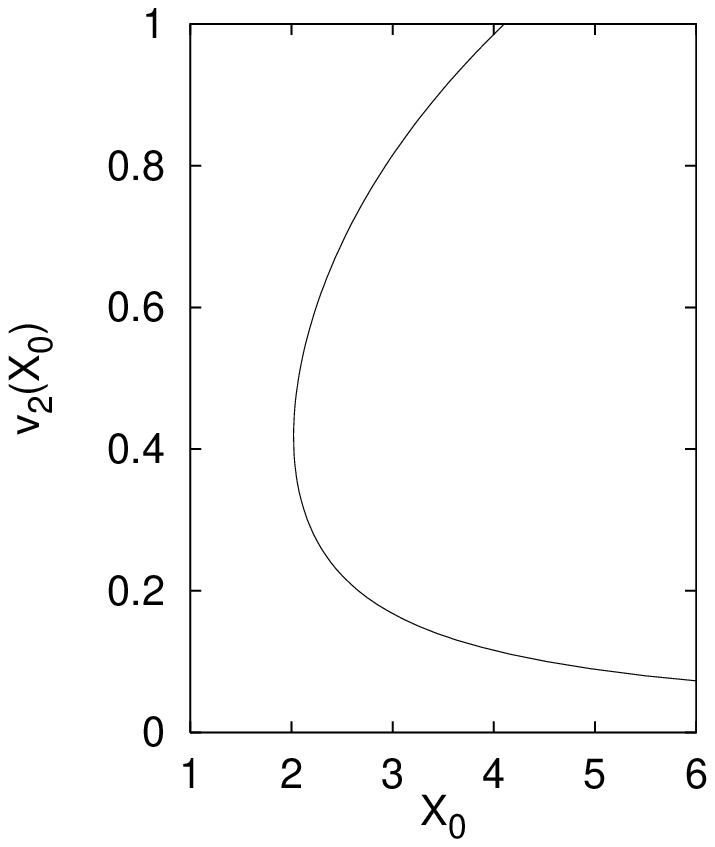}{100mm}{1}{} 
\pagebreak[5]
\ifig{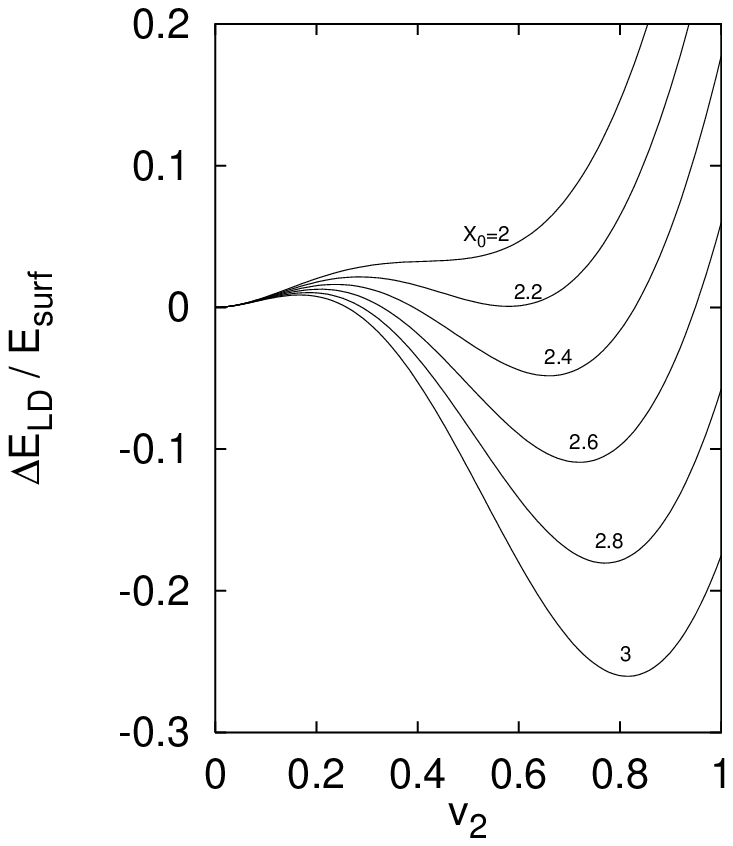}{100mm}{2}{}
\pagebreak[5]
\ifig{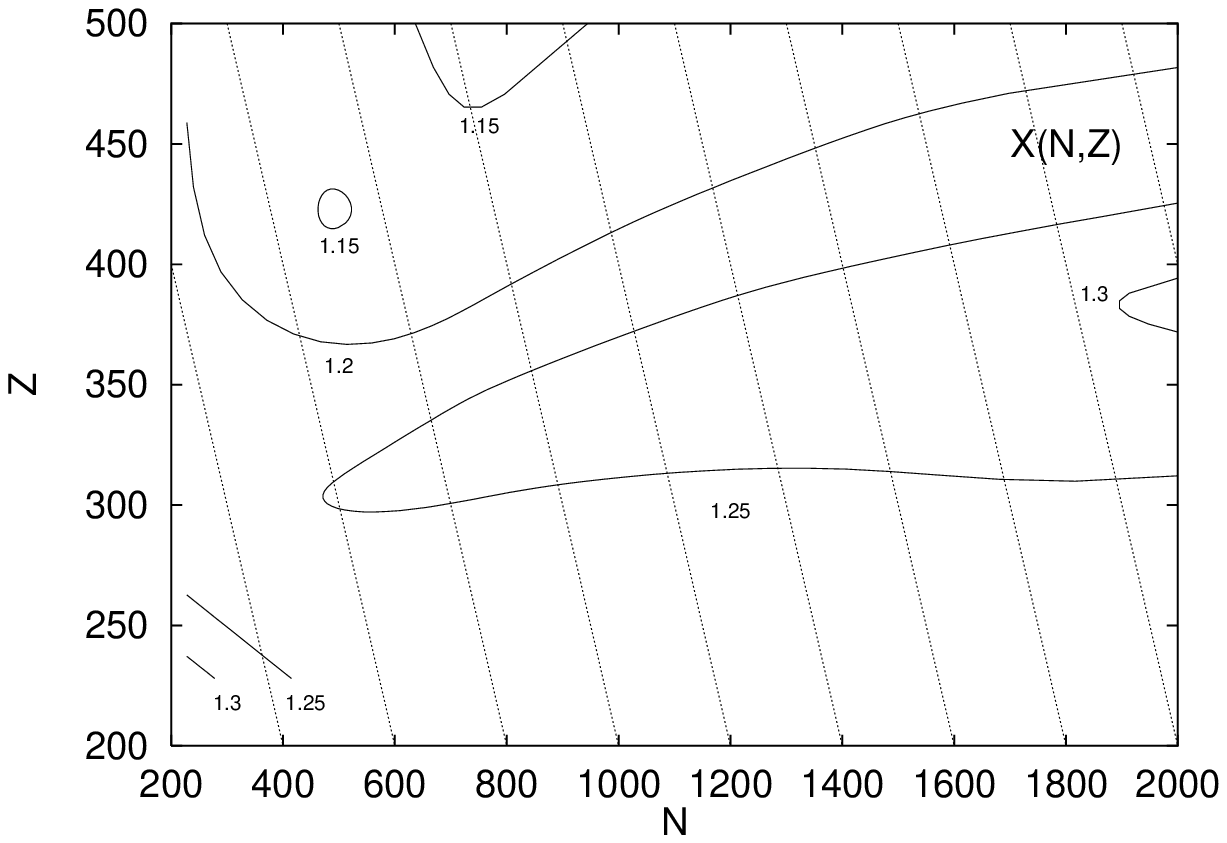}{100mm}{3}{}
\pagebreak[5]
\ifig{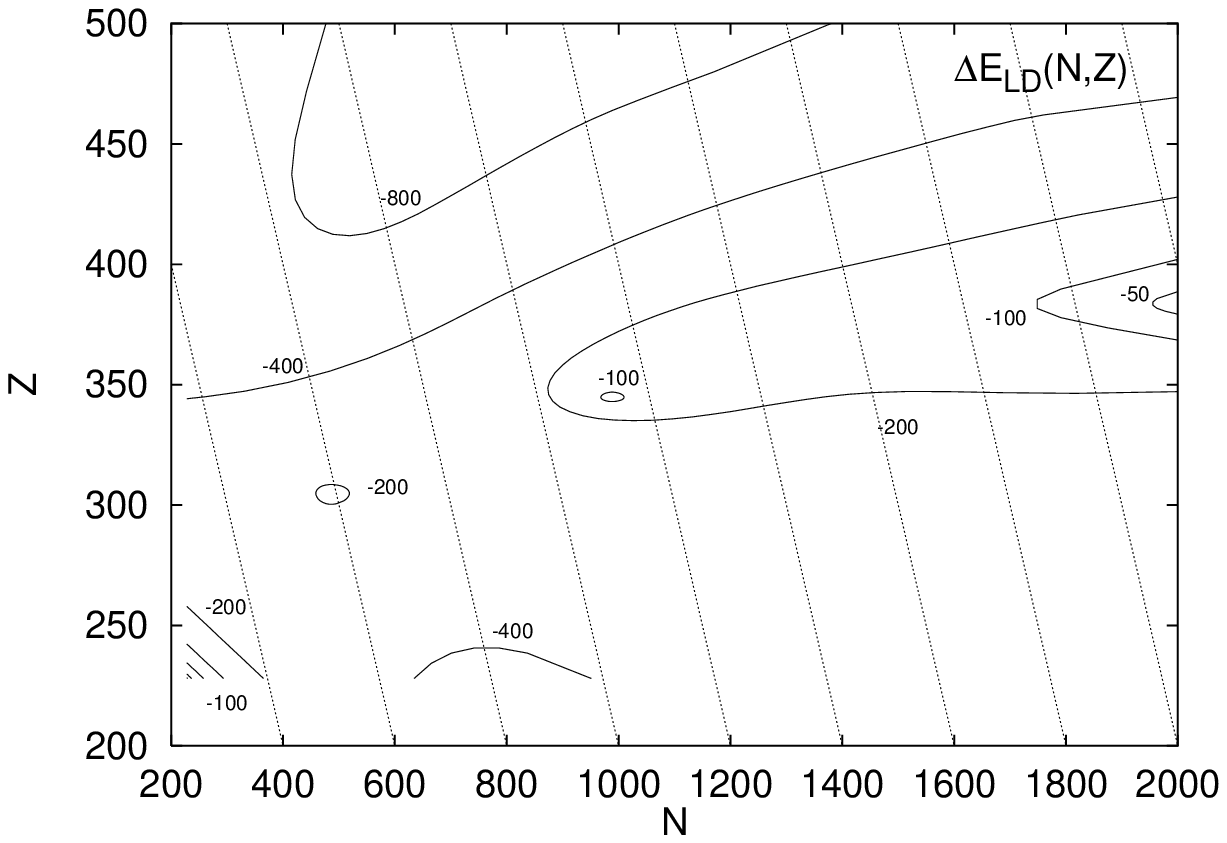}{100mm}{4}{}
\pagebreak[5]
\ifig{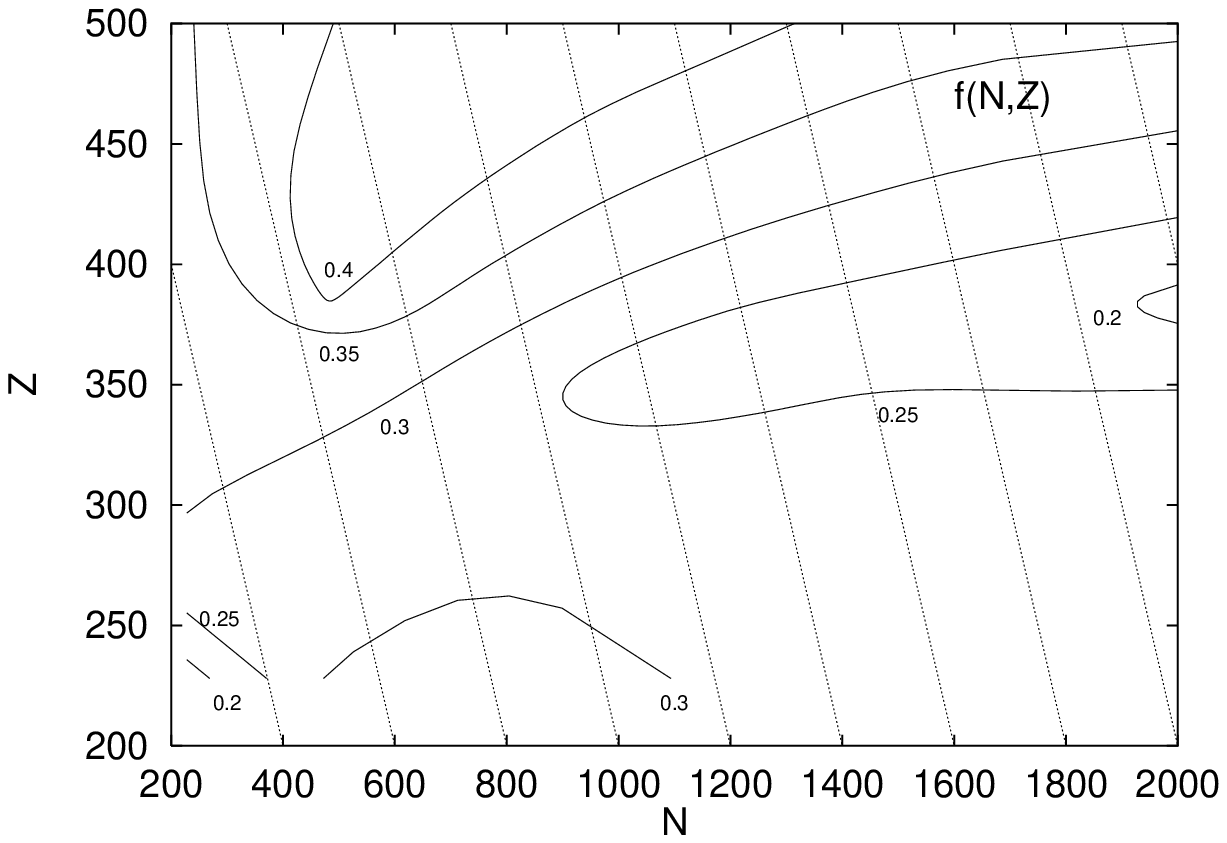}{100mm}{5}{}
\pagebreak[5]
\ifig{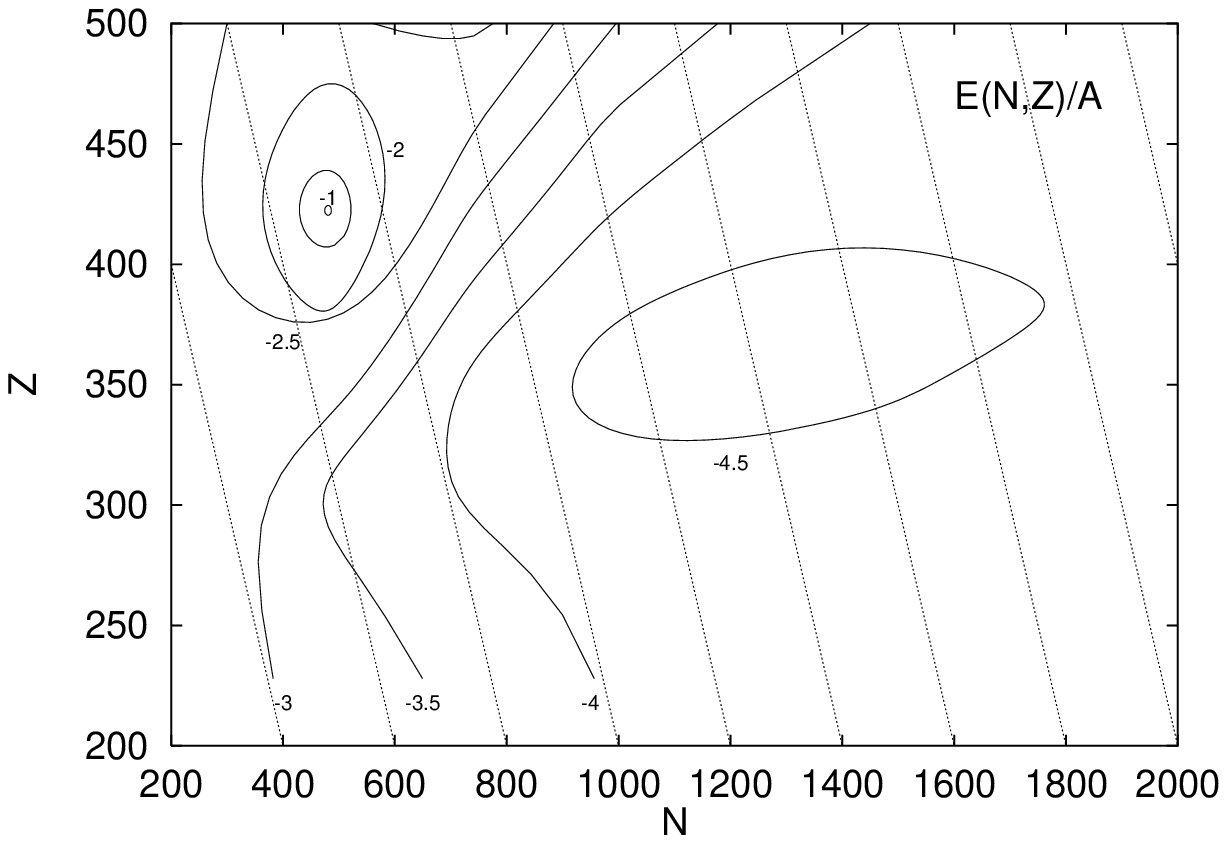}{100mm}{6}{}
\pagebreak[5]
\ifig{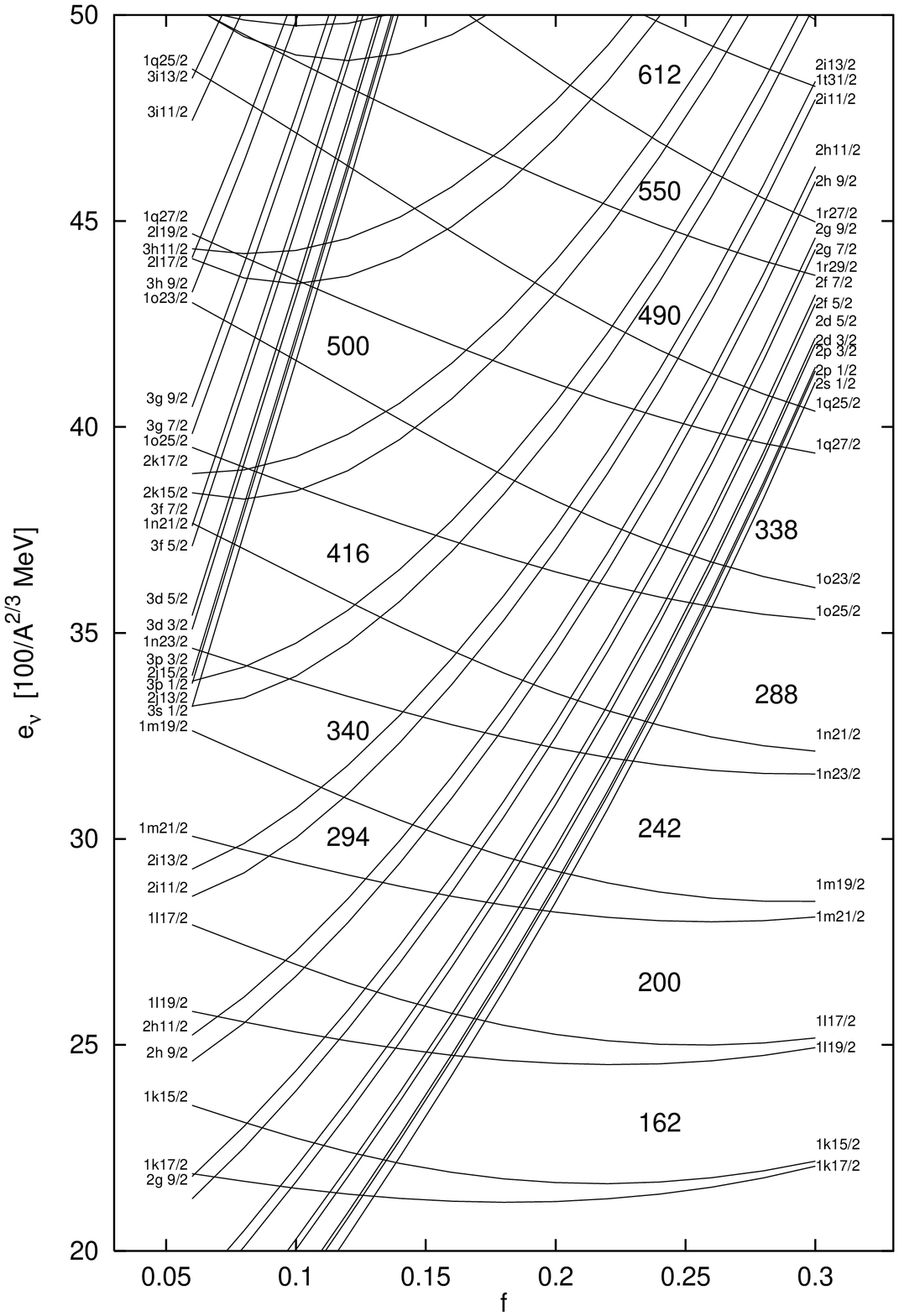}{180mm}{7}{}
\pagebreak[5]
\ifig{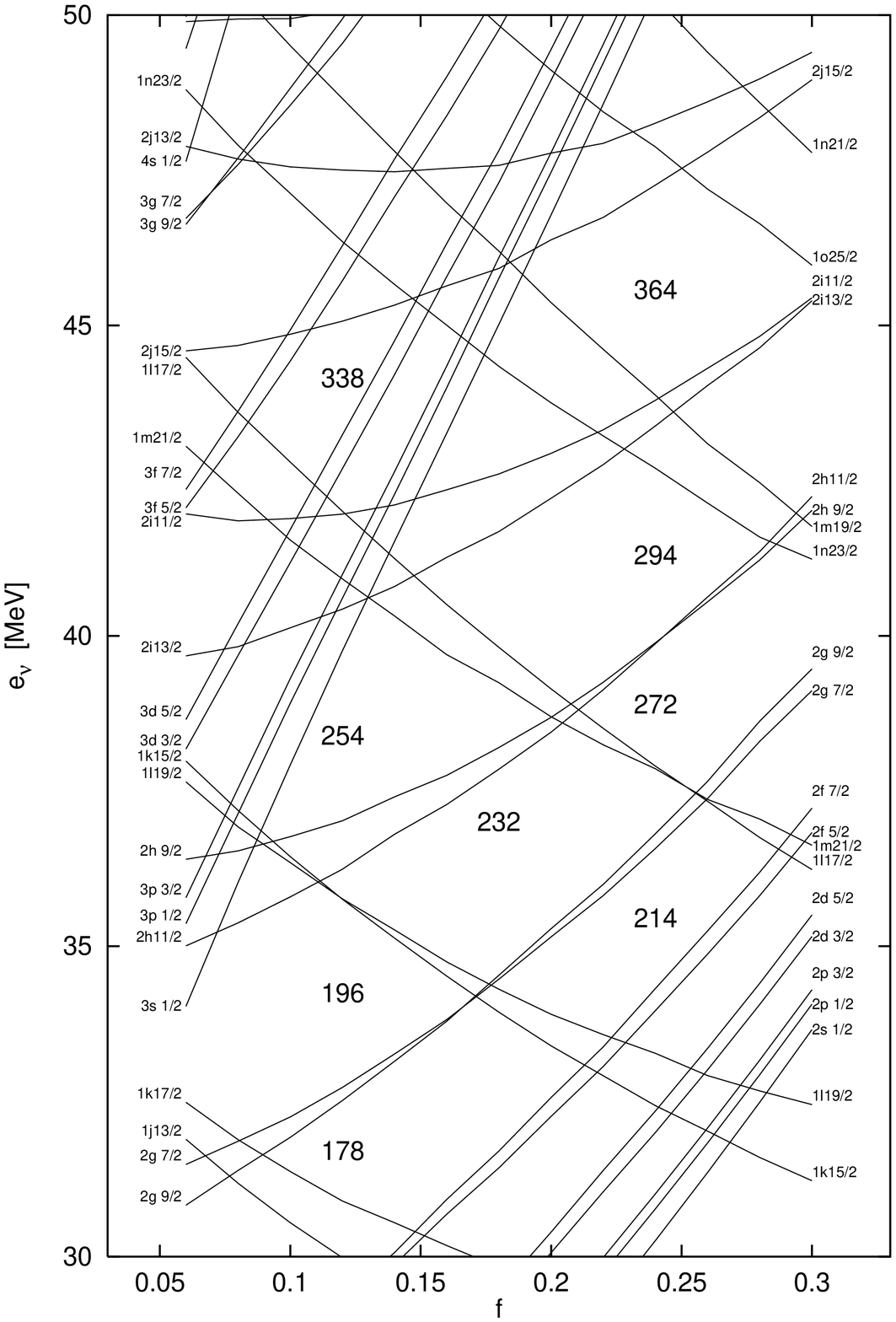}{180mm}{8}{}
\pagebreak[5]
\ifig{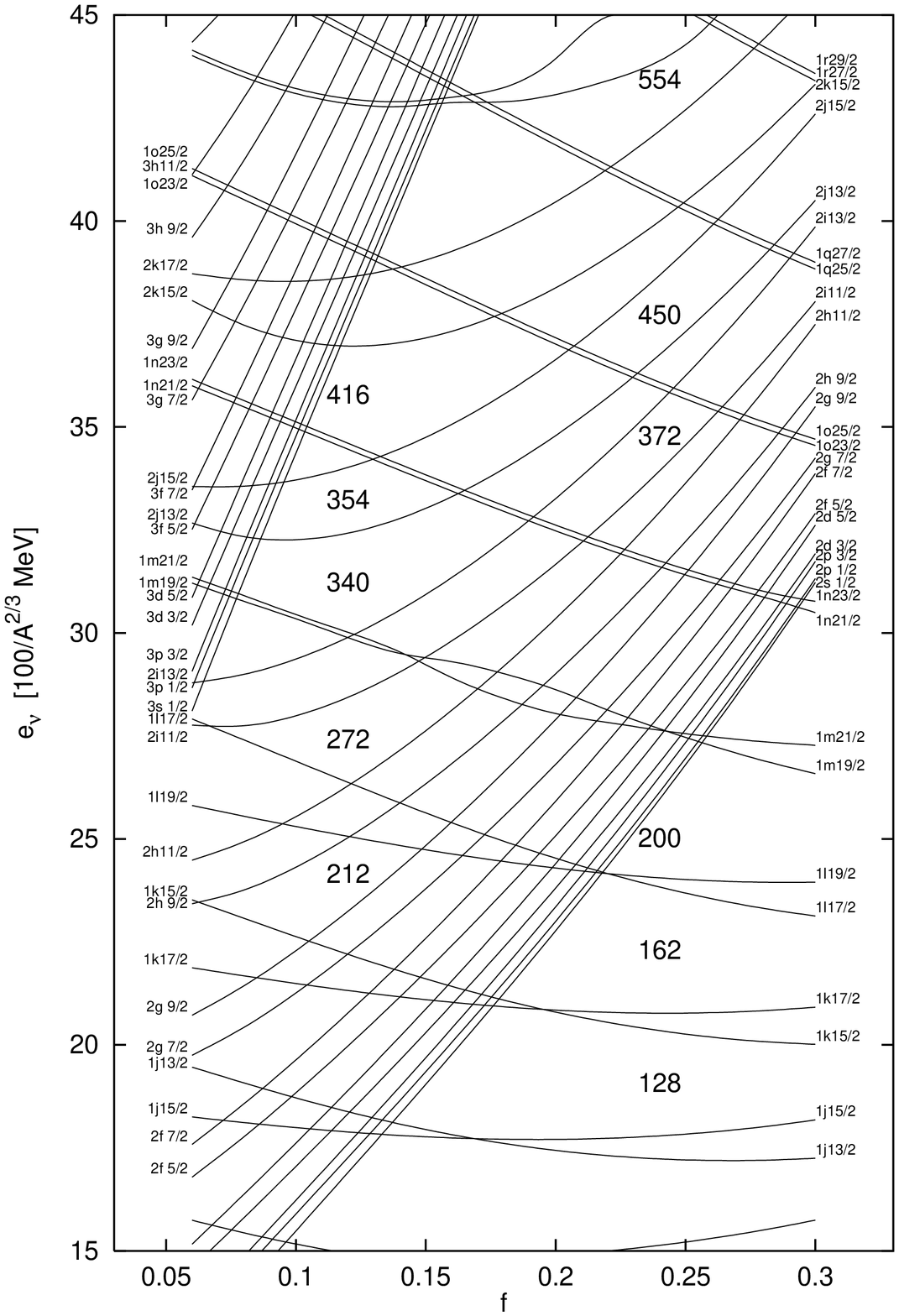}{180mm}{9}{}
\pagebreak[5]
\ifig{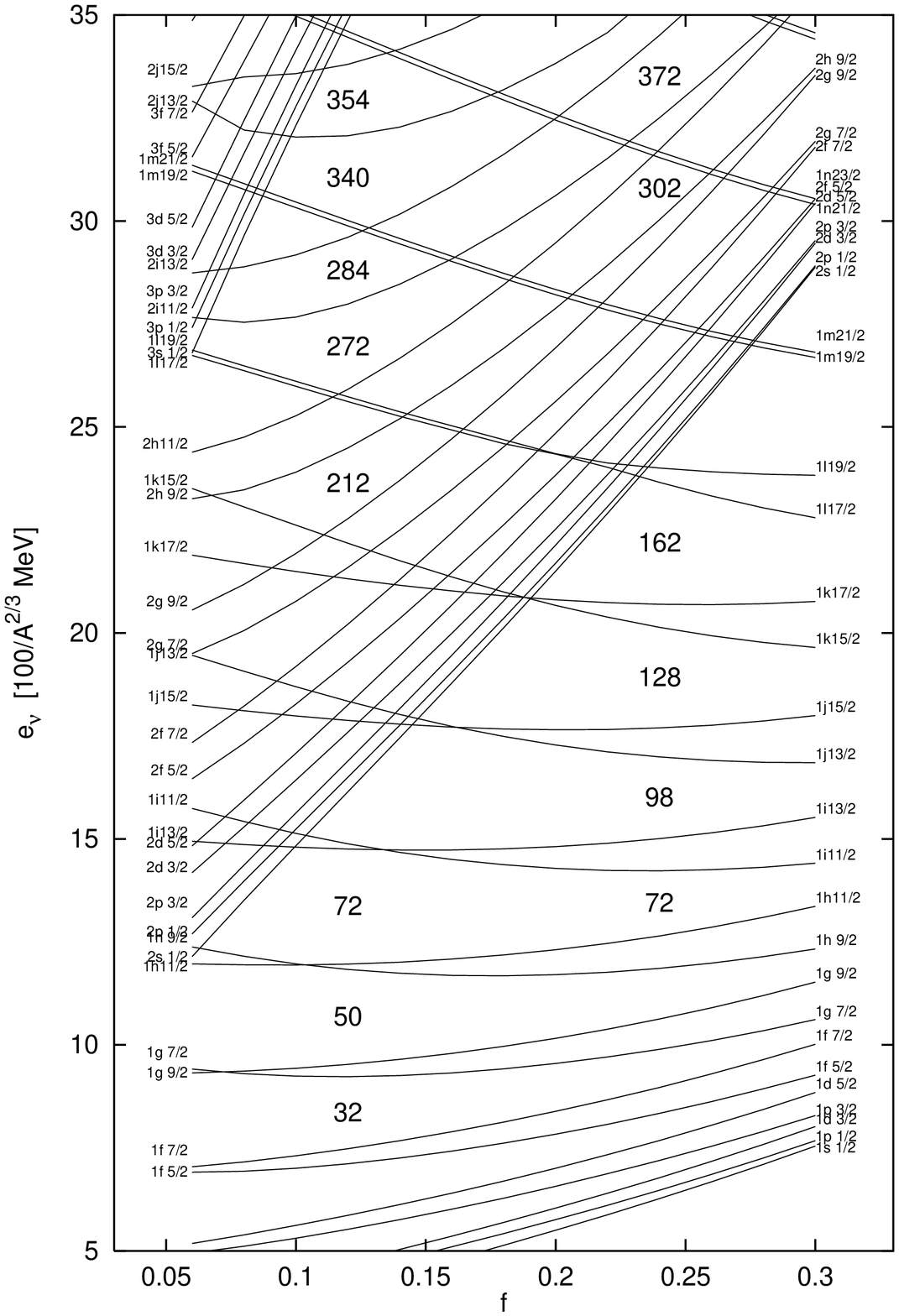}{180mm}{10}{}
\pagebreak[5]
\ifig{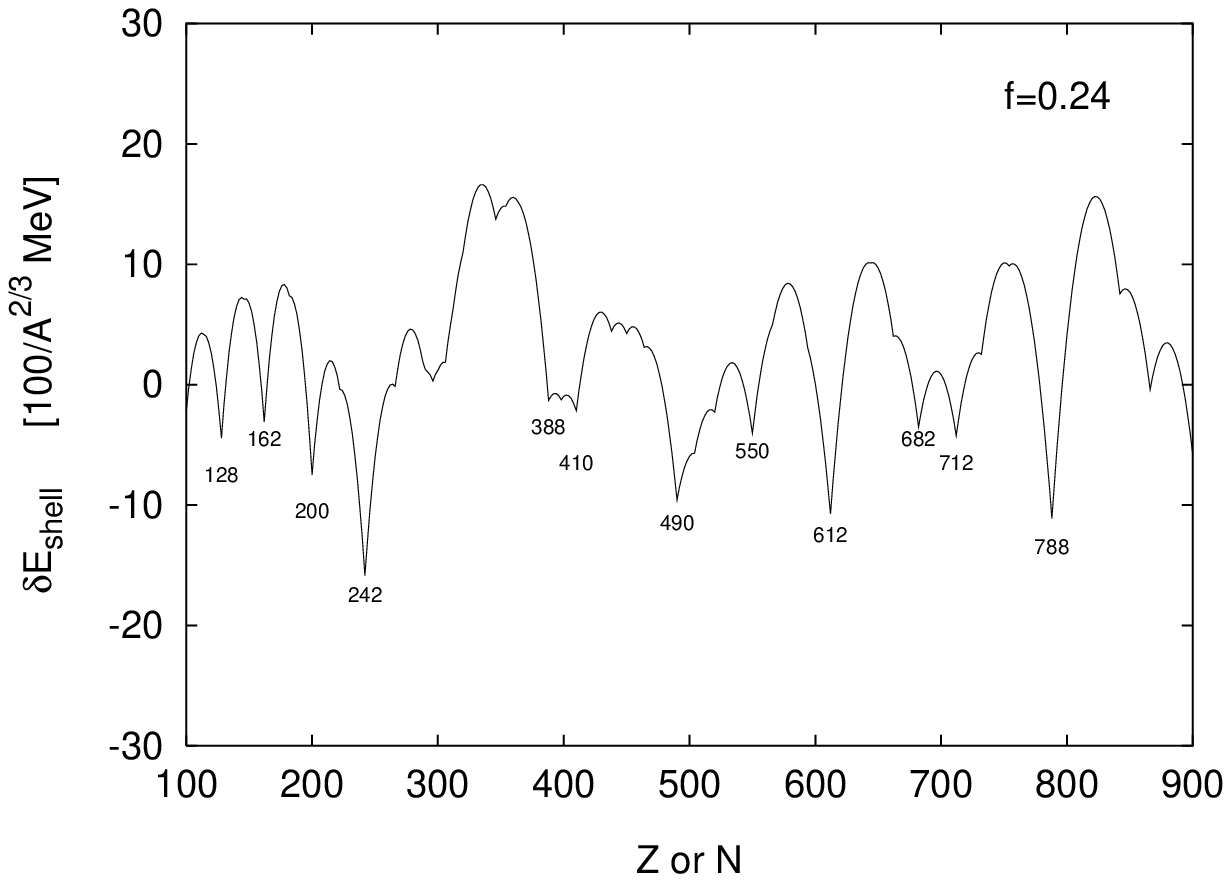}{100mm}{11}{}
\pagebreak[5]
\ifig{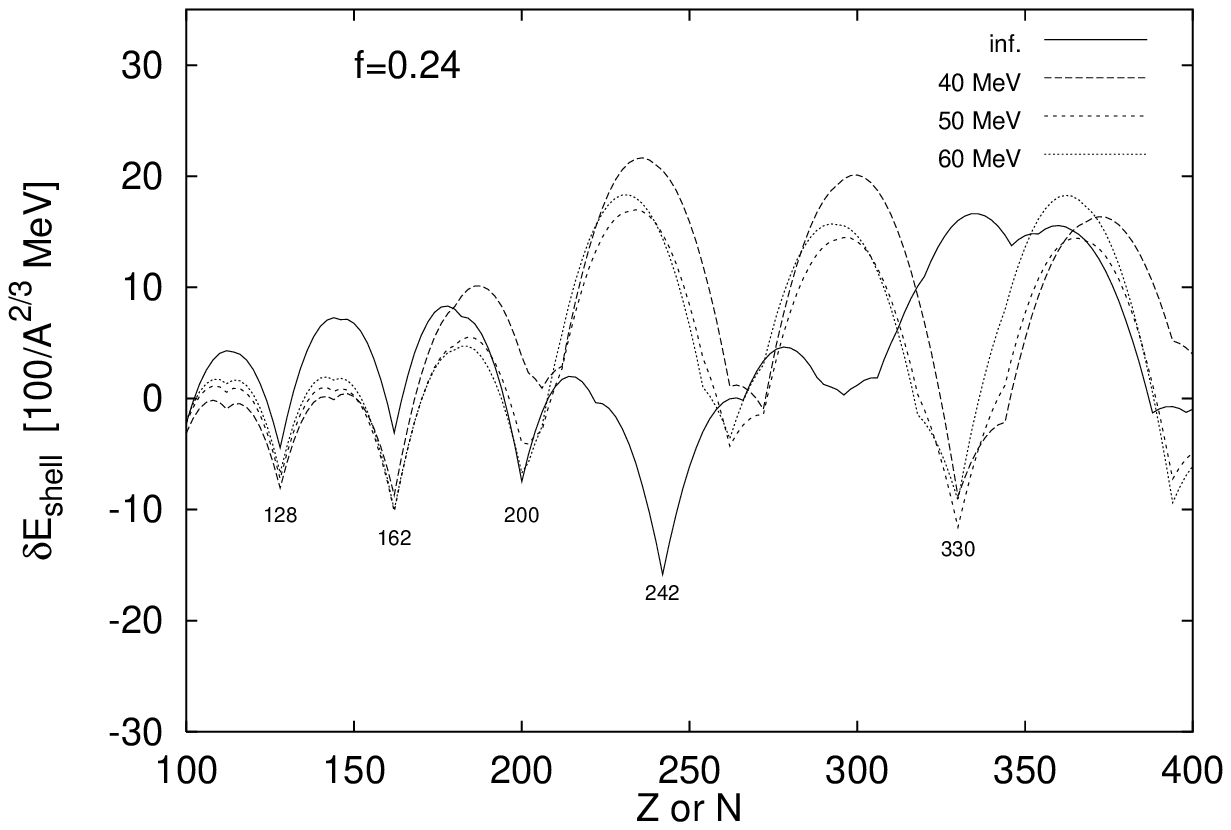}{100mm}{12}{}
\pagebreak[5]
\ifig{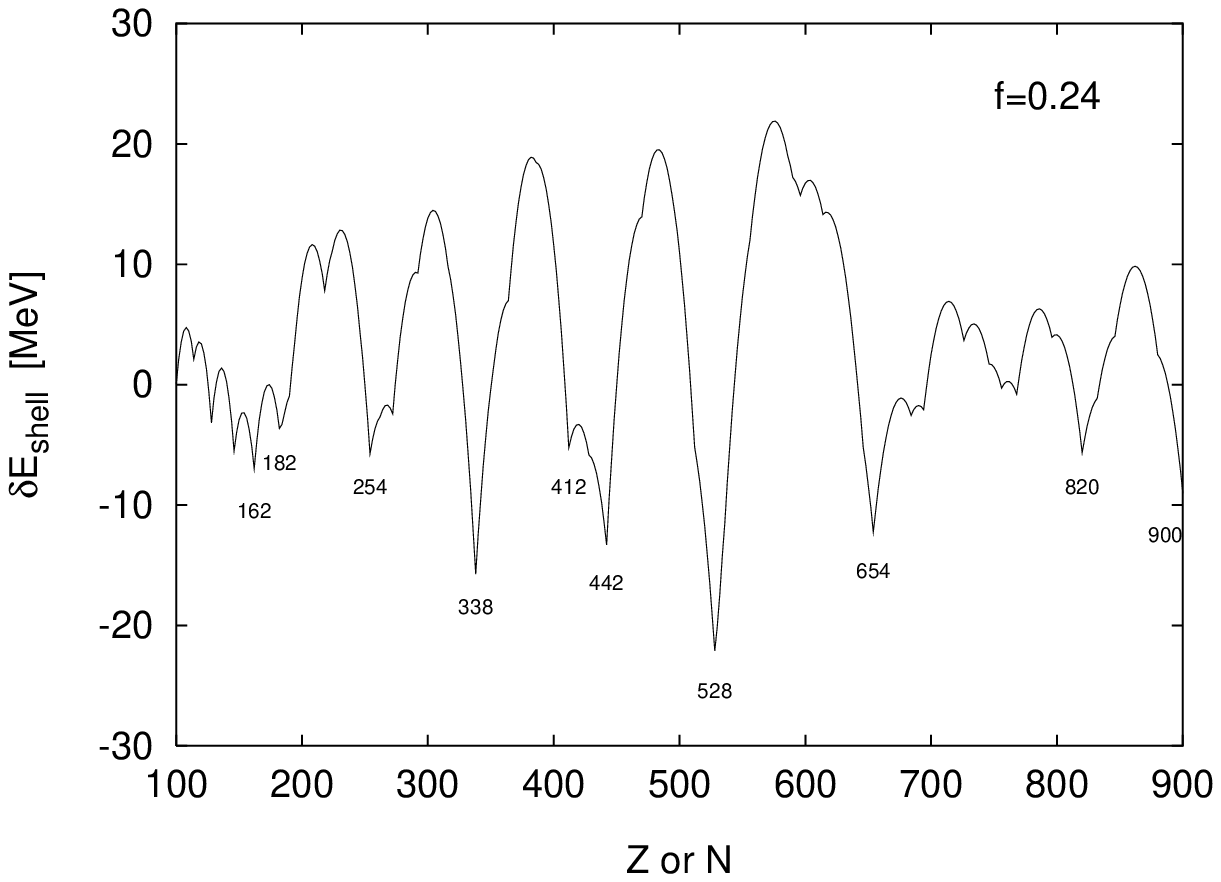}{100mm}{13}{}
\pagebreak[5]
\ifig{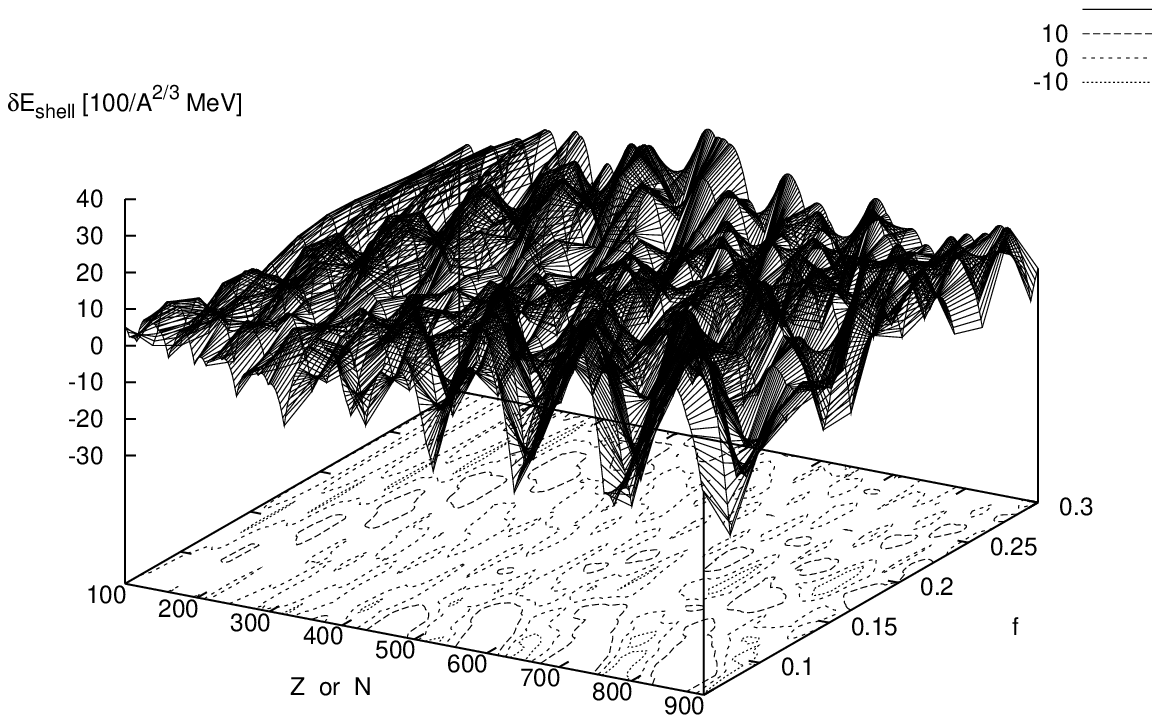}{100mm}{14}{}
\pagebreak[5]
\ifig{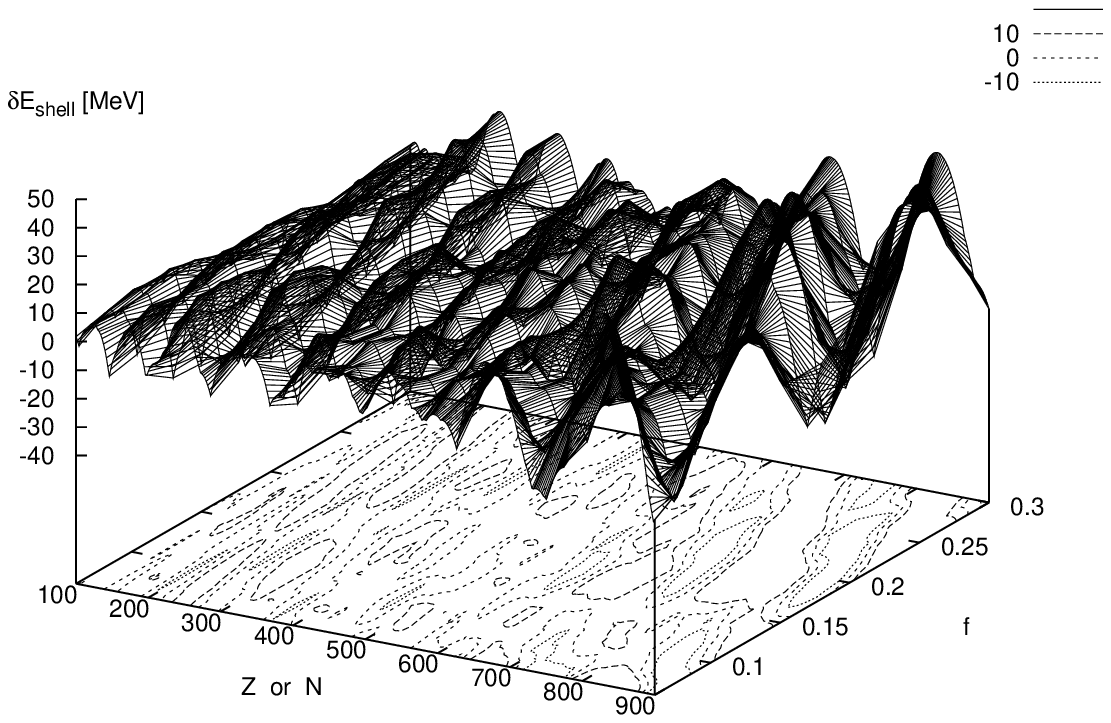}{100mm}{15}{}
\pagebreak[5]
\ifig{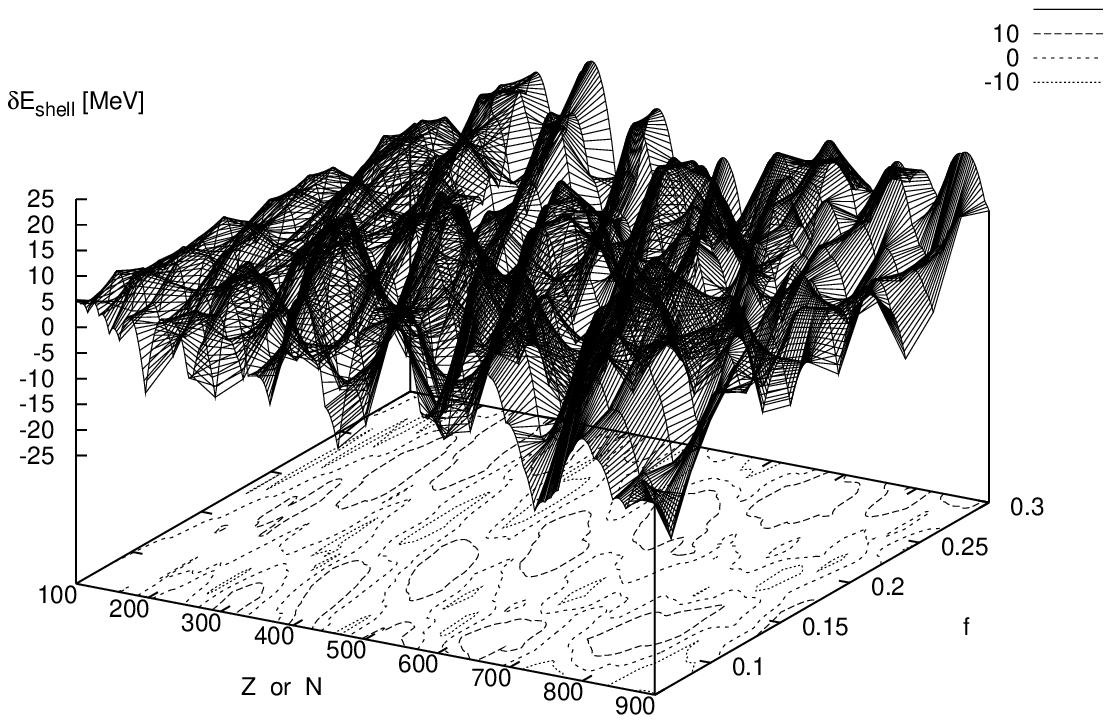}{100mm}{16}{}
\pagebreak[5]
\ifig{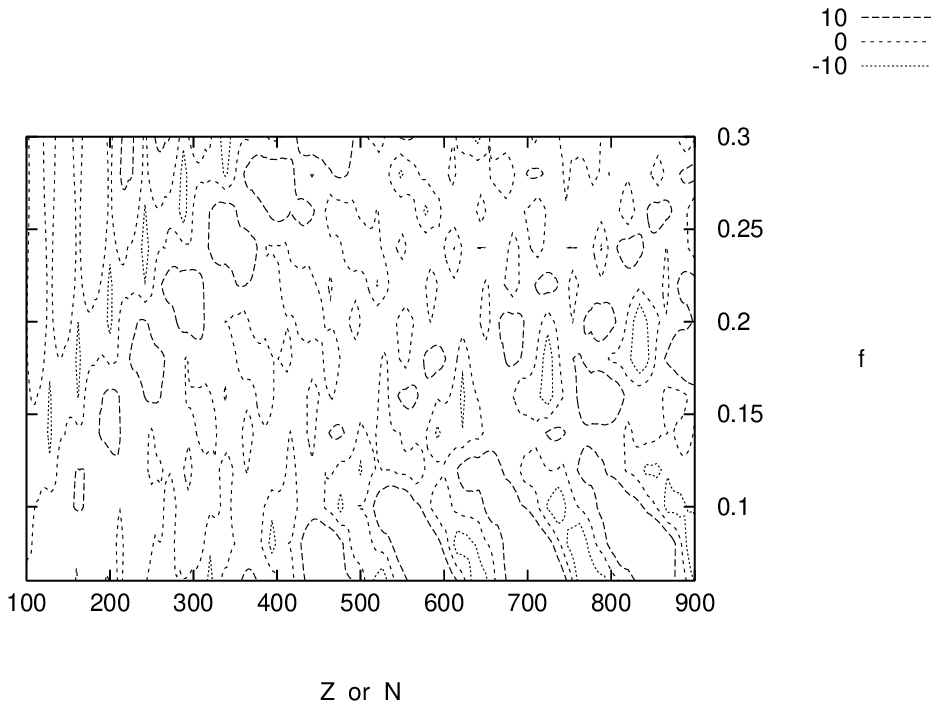}{100mm}{17}{}
\pagebreak[5]
\ifig{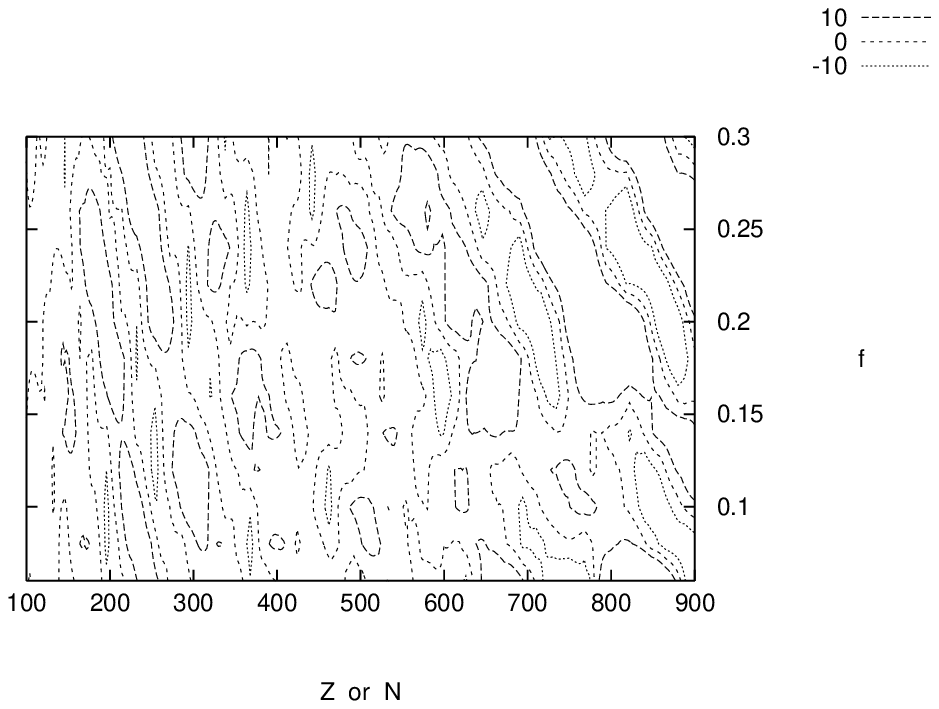}{100mm}{18}{}
\pagebreak[5]
\ifig{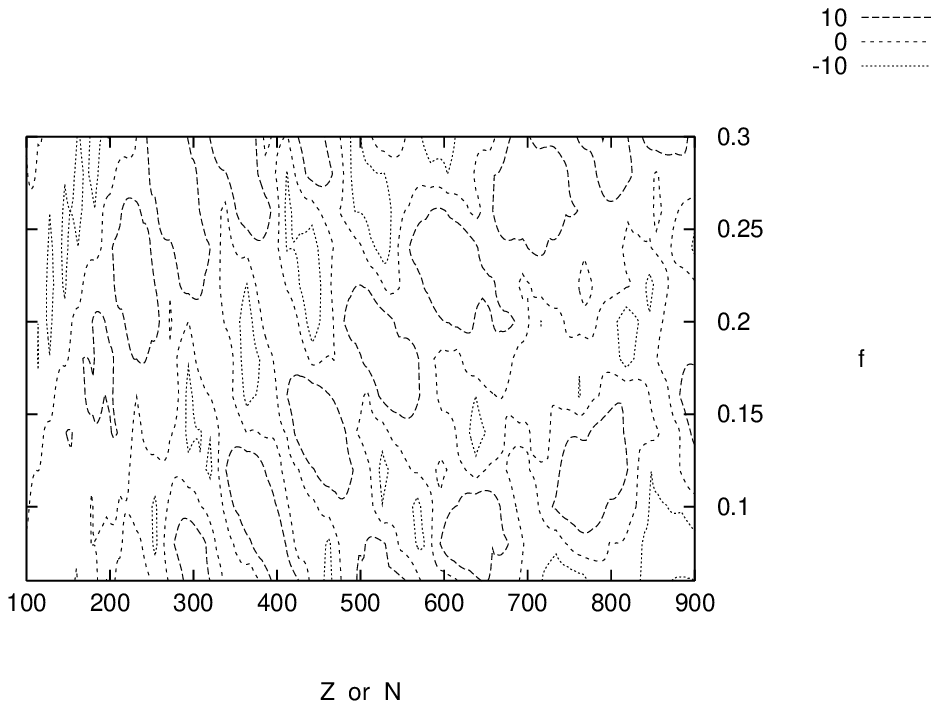}{100mm}{19}{}
\pagebreak[5]
\ifig{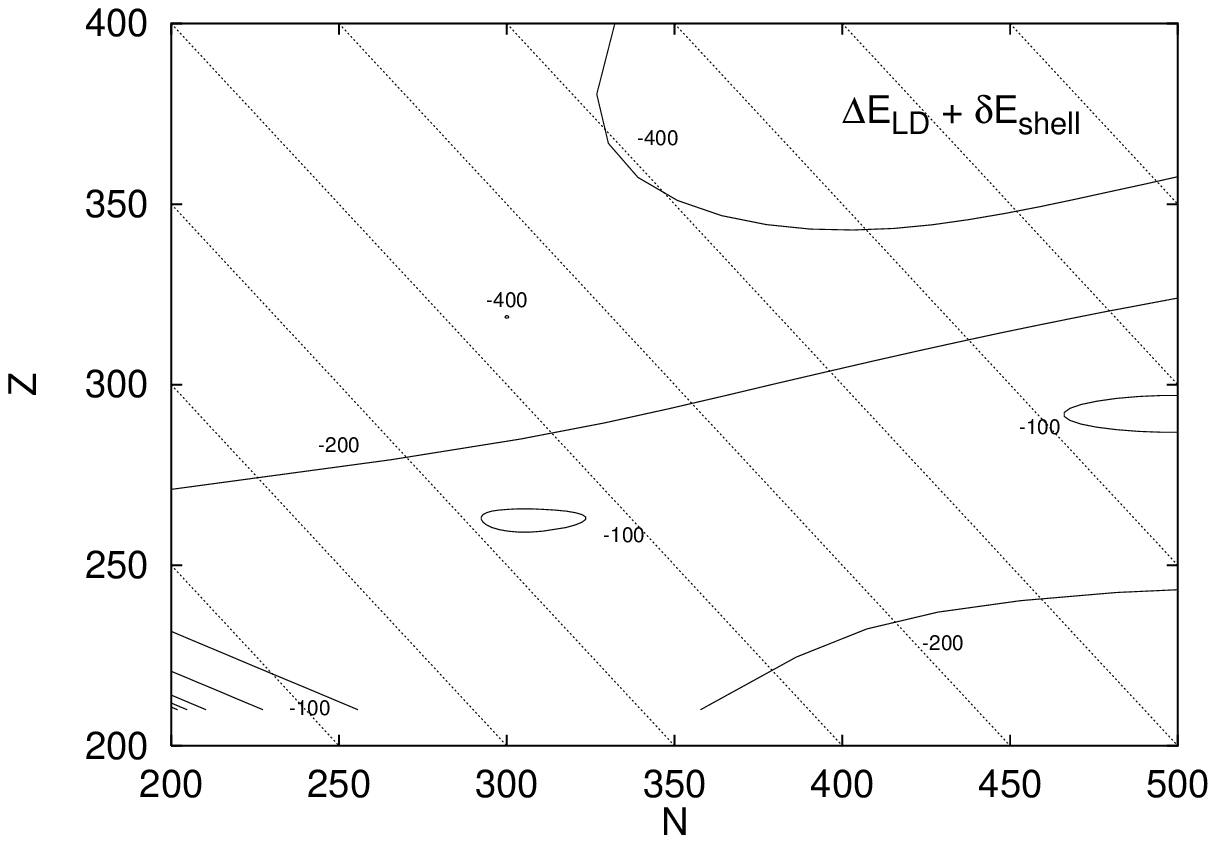}{100mm}{20}{}
\pagebreak[5]
\ifig{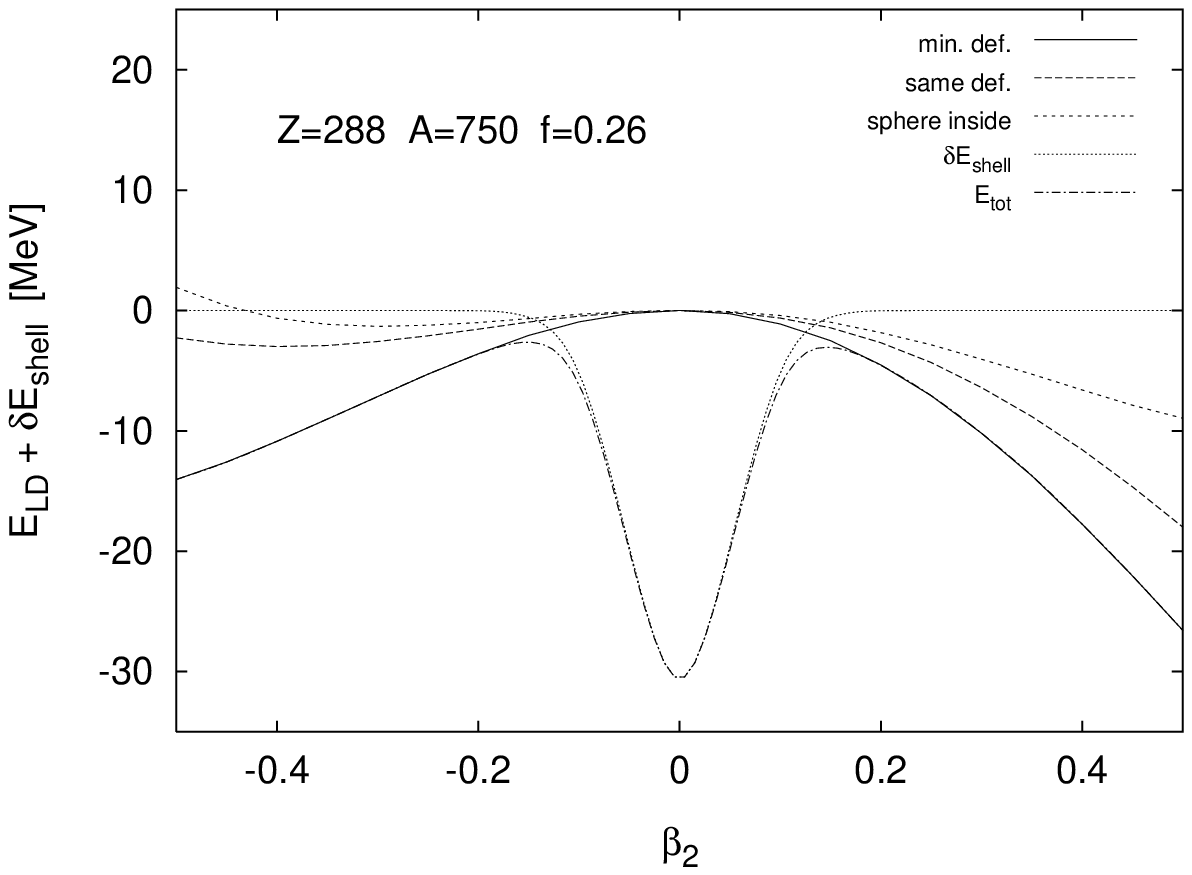}{100mm}{21}{}
\pagebreak[5]
\ifig{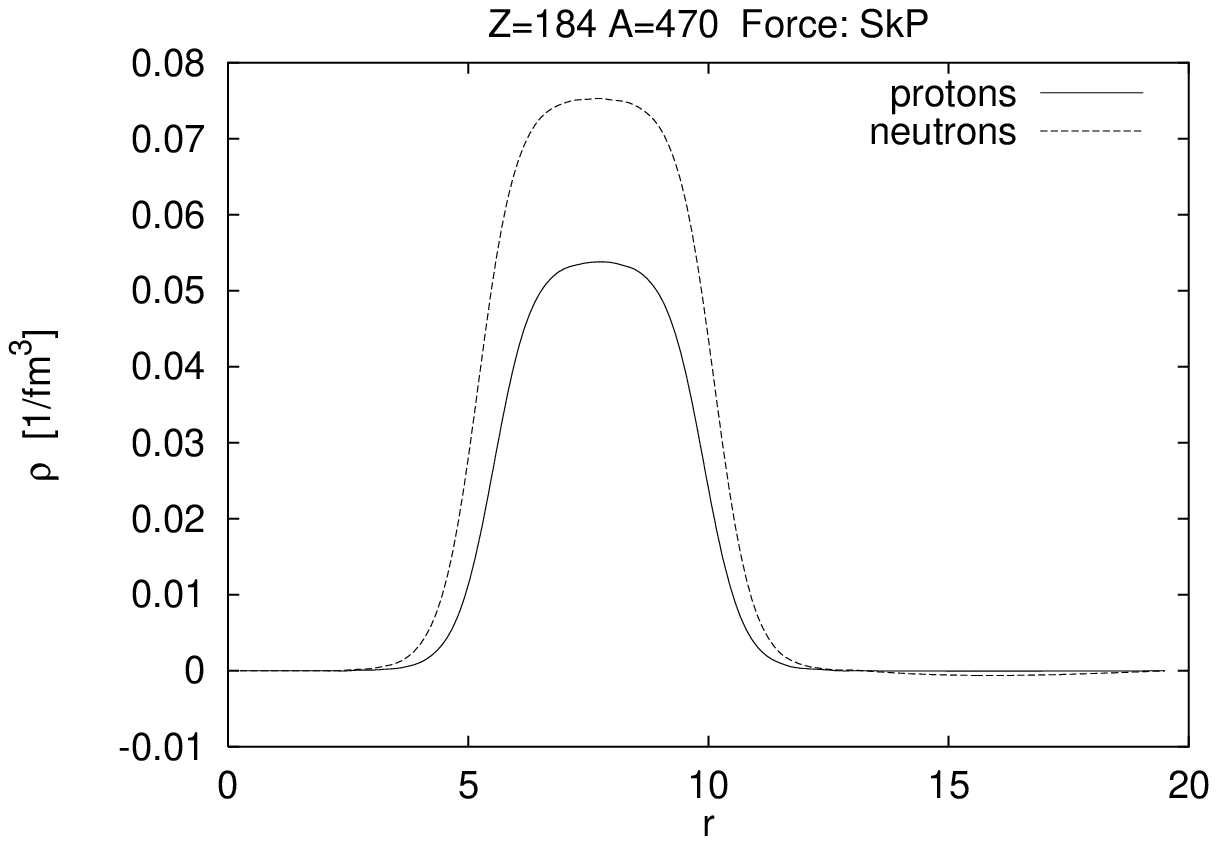}{100mm}{22}{}
\pagebreak[5]
\ifig{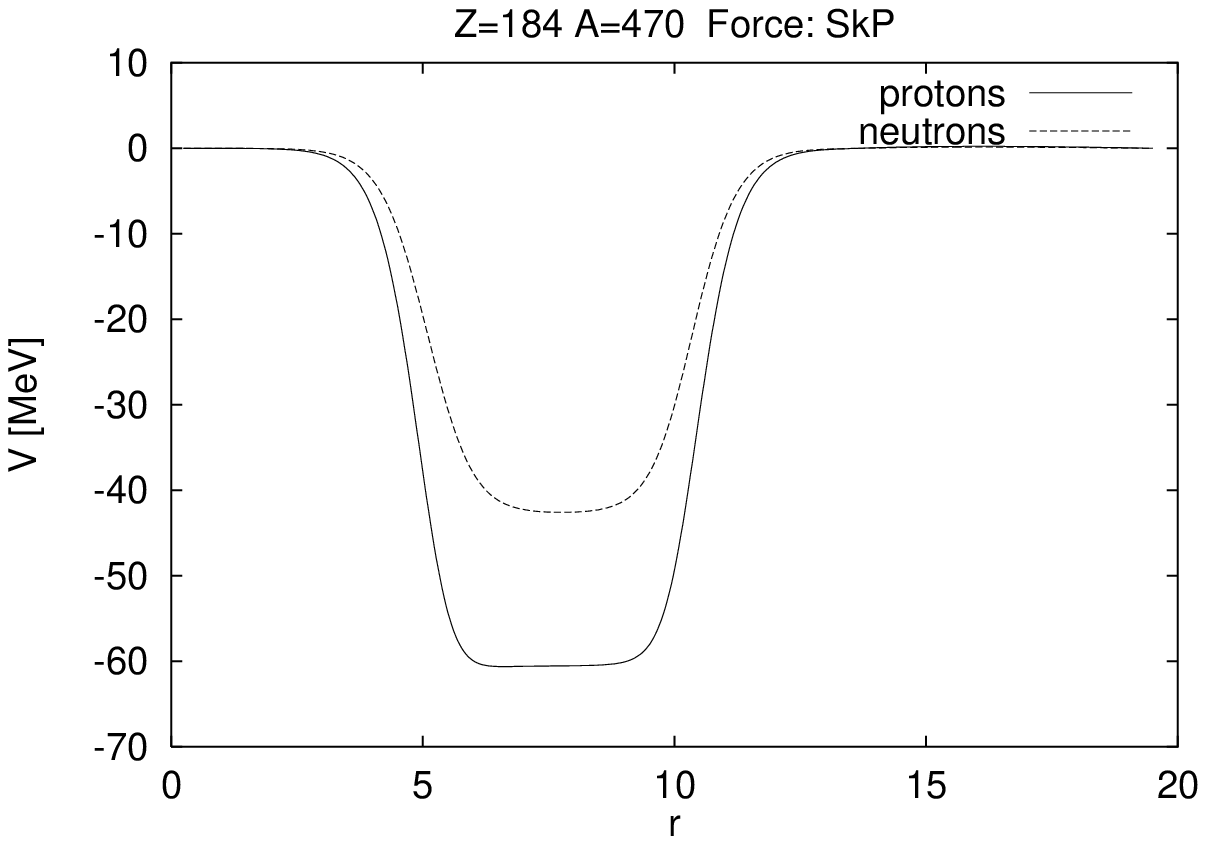}{100mm}{23}{}
\pagebreak[5]
\ifig{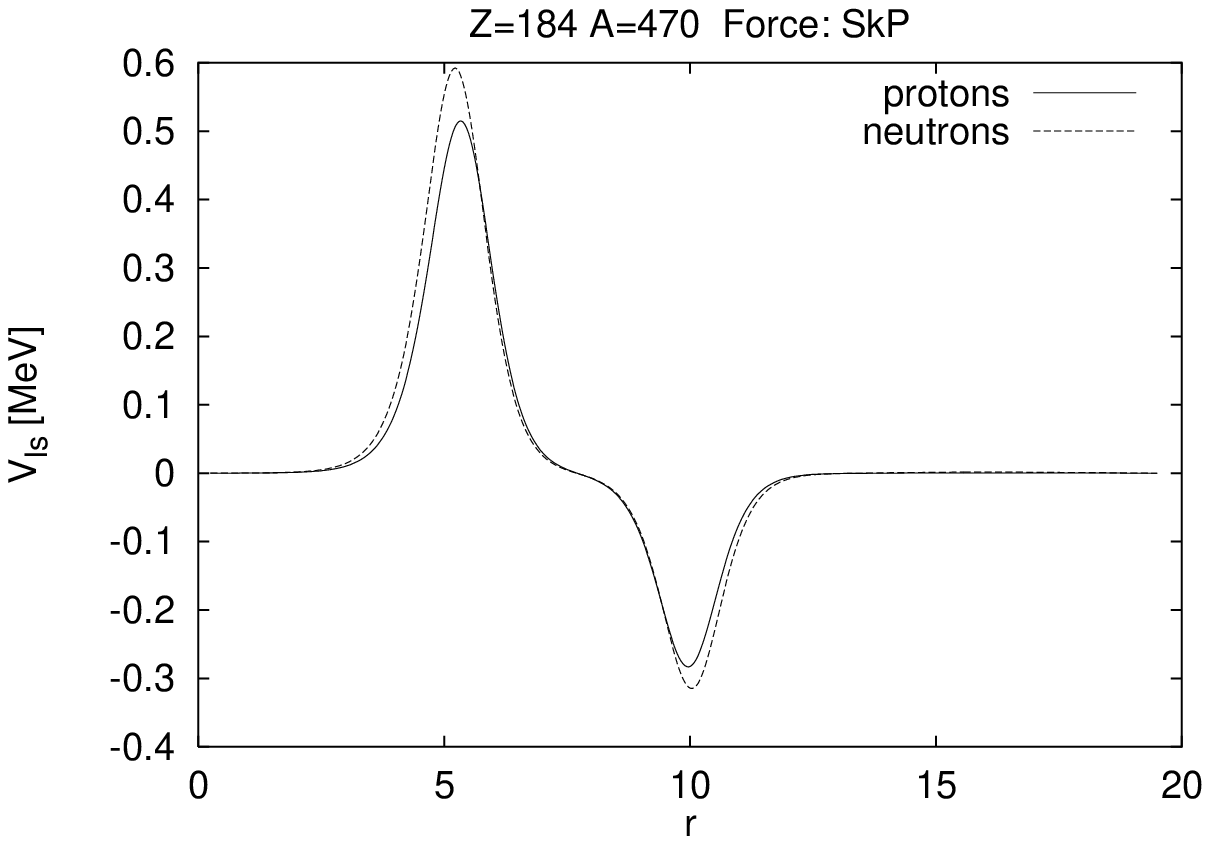}{100mm}{24}{}
\clearpage
\ifig{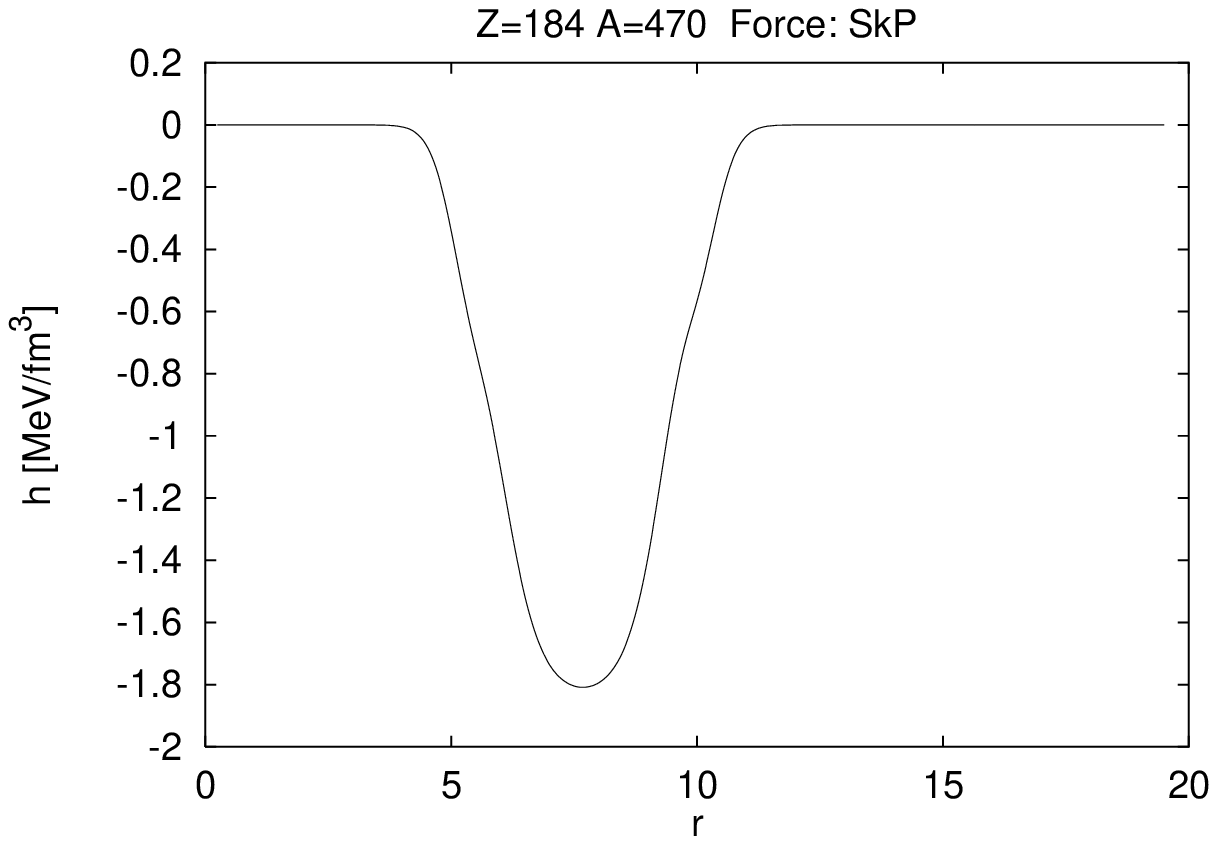}{100mm}{25}{}
\pagebreak[5]
\ifig{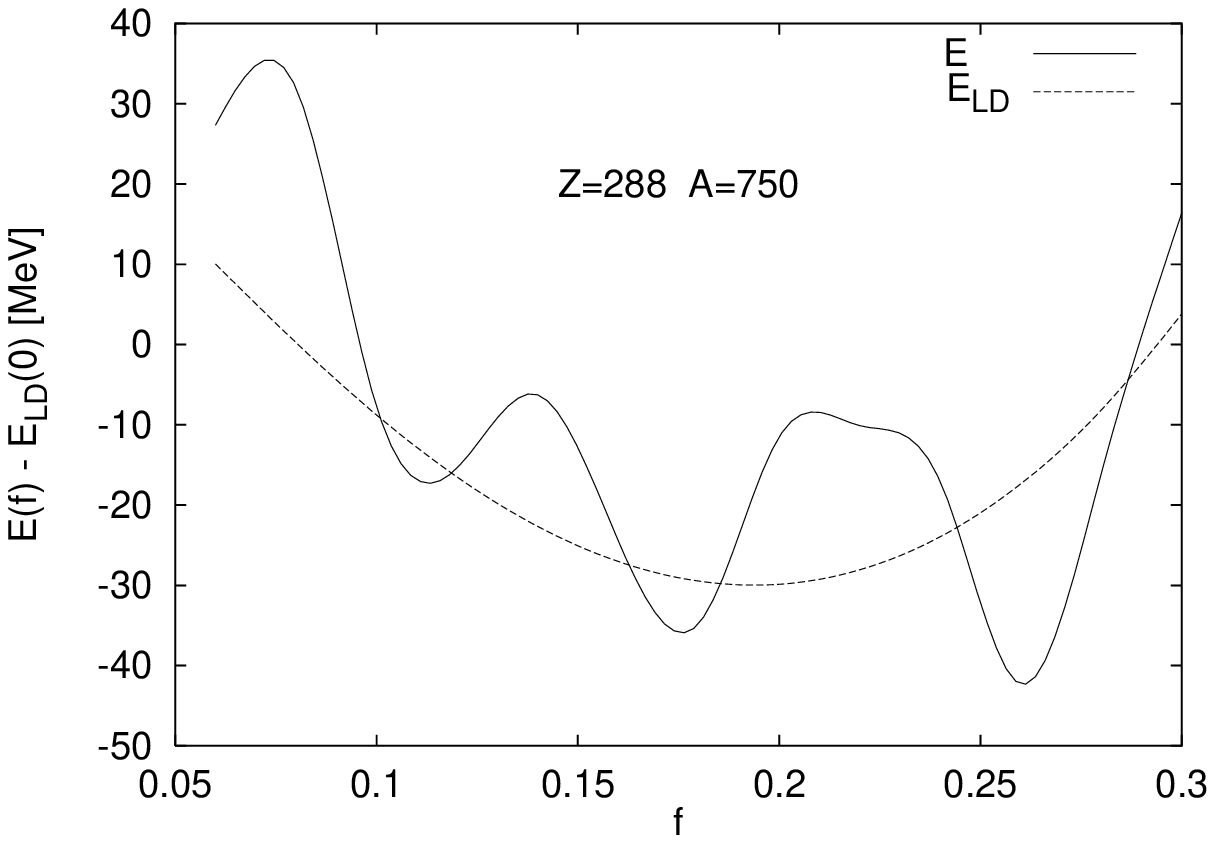}{100mm}{24}{}
\pagebreak[5]
\ifig{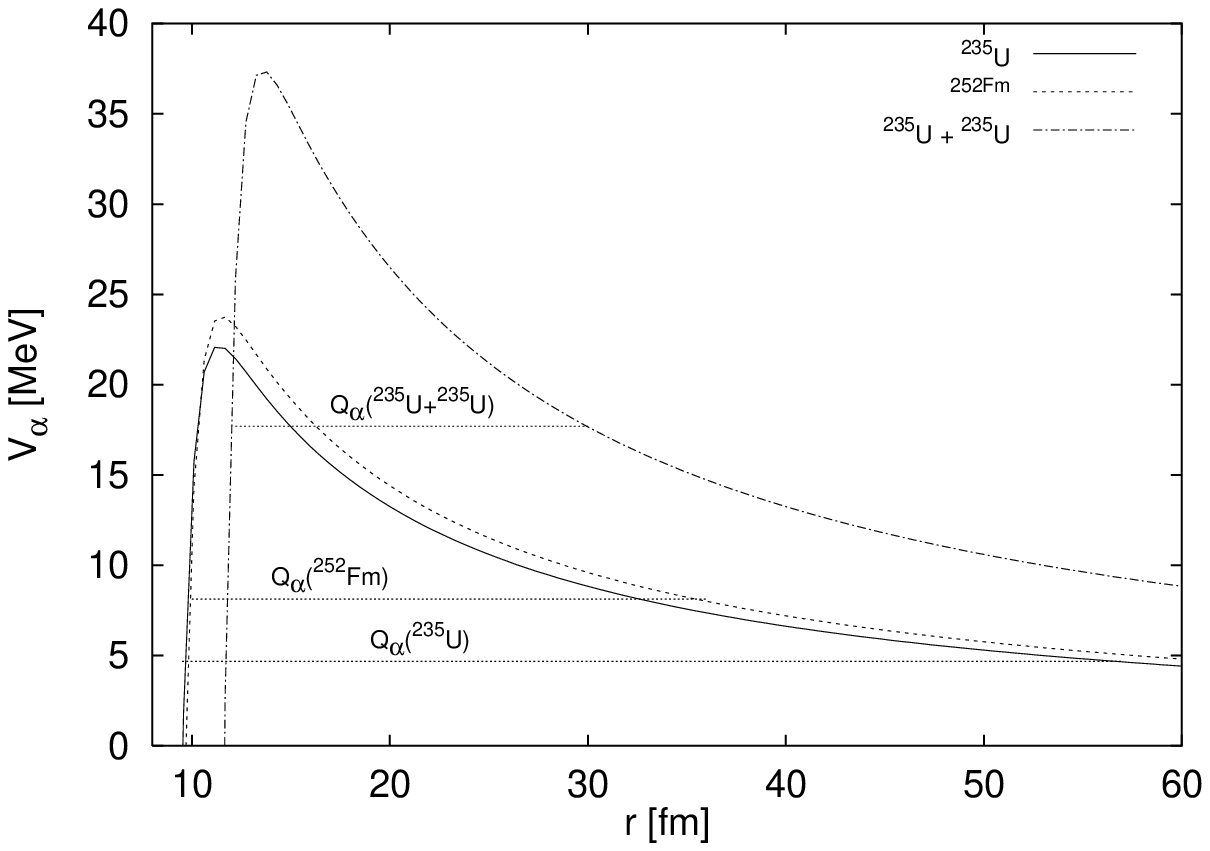}{100mm}{25}{}

\end{document}